\documentclass[a4,fproof]{cta-author}

{}
{}
{}
\usepackage{amsmath}
\usepackage{amsfonts}
\newcommand{\Z}{\mathbb{Z}}
\usepackage[caption=false,font=footnotesize]{subfig}
\usepackage{mathtools}
\usepackage{nccmath}
\usepackage{multirow}
\usepackage{graphicx} % for '\resizebox` macro
\usepackage{tabularx} % for 'tabularx' environment
\usepackage{enumerate}   

\begin{document}

%\supertitle{Submission Template for IET Research Journal Papers}

\title{Analysis of power system inertia estimation in high wind power plant integration scenarios}

\author{\au{Ana~Fern\'{a}ndez-Guillam\'{o}n$^{1}$}, \au{Antonio~Vigueras-Rodr\'{i}guez$^{2}$}, \au{\'{A}ngel~Molina-Garc\'{i}a$^{1\corr}$}}

\address{\add{1}{Department of Automatics, Electrical Eng. and Electronic Technology, Universidad Politecnica de Cartagena, 30202 Cartagena, Spain}
\add{2}{Department of Civil Engineering, Universidad Politecnica de Cartagena, 30203 Cartagena, Spain}
\email{angel.molina@upct.es}}

\begin{abstract}
{\color{black}{Nowadays, power system inertia is changing as a consequence of replacing conventional units by renewable energy sources, mainly wind and PV power plants.}} This fact affects significantly the grid frequency response under power imbalances. As a result, new frequency control strategies for renewable plants are being developed to emulate the behaviour of conventional power plants under such contingencies. These approaches are usually called 'virtual inertia emulation techniques'. {\color{black}{In this paper, an analysis of power system inertia estimation from frequency excursions is carried out by considering different inertia estimation methodologies, discussing the applicability and coherence of these methodologies under the new supply-side circumstances.}} The modelled power system involves conventional units and wind power plants, including wind frequency control strategies in line with current mix generation scenarios. %Therefore, the virtual inertia provided by wind power plants can be estimated. 
Results show that all methodologies considered provide an accurate result to estimate the equivalent inertia based on rotational generation units directly connected to the grid. However, %most methodologies present 
significant discrepancies are found when frequency control strategies are included in wind power plants decoupled from the grid. In this way, authors consider that it is necessary to define alternative inertia estimation methodologies by including virtual inertia emulation. Extensive discussion and results are also provided in this study.    

\end{abstract}

\maketitle

\section{Introduction}\label{sec.introduction}

Frequency of a power system deviates from its nominal value after a severe power imbalance between generation and consumption~\cite{babahajiani18}. Due to the increasing penetration of renewable energy sources (RES), mainly wind and PV, electrical grids can suffer more frequency stability challenges~\cite{cvetkovic17}. RES are intermittent and uncertain because they depend on weather conditions~\cite{wang15}. This fact makes them hard to integrate into power systems~\cite{teng17}, as they pose stress on their operation~\cite{rodriguez14}: Transmission System Operators (TSOs) have to deal with not only the uncontrollable demand but also uncontrollable generation~\cite{zhang17}.

Moreover, renewable power plants are not connected to the grid through synchronous machines, but through electronic converters~\cite{junyent15}. Thus, by increasing the amount of renewable sources and replacing synchronous conventional units, the effective rotational inertia of the system can be significantly reduced~\cite{akhtar15,yang18}. The rotational inertia is important to limit the rate of change of frequency (ROCOF) right after a power imbalance~\cite{dehghanpour15}. Therefore, power systems with lower equivalent inertia are initially more sensitive to frequency deviations~\cite{kim16,nguyen18}. As a result, frequency control strategies have been developed to effectively integrate RES into the grid~\cite{gross17increasing}. Such methods are 
commonly referred to as synthetic, artificial, emulated or virtual inertia~\cite{vokony17}. 

The aim of this paper is to estimate and compare the equivalent inertia constant of a power system with high RES integration from the frequency deviations suffered after an imbalance. Several methodologies have been proposed during the last decades in the specific literature~\cite{inoue97,chassin05,wall12,wall14,zografos17,tuttelberg18,zografos18}. The power system considered in this paper is in line with current grids, involving conventional and wind power plants. Moreover, wind plants include frequency control according to a recent approach~\cite{fernandez18}. The rest of the paper is organized as follows: the theoretical background of the problem is covered in Section~\ref{sec.background}. Section~\ref{sec.inertia_estimation} reviews and explains the different strategies to estimate the inertia constant of a power system after an imbalance. In Section~\ref{sec.system_identification}, the power system and different scenarios considered in this paper are detailed. Results are discussed in Section~\ref{sec.results}. Conclusions are given in Section~\ref{sec.conclusion}.

\section{Theoretical background}\label{sec.background}
\subsection{Inertia constant $H$} \label{sec.inertia_constant}
From a traditional point of view, after a power imbalance, the kinetic energy stored in the rotating masses of a generator is released following expression~\eqref{eq.Ekin}~\cite{ulbig14}:
\begin{align}
\label{eq.Ekin}
E_{kin}=\dfrac{1}{2}\;J\;(2\cdot\pi\cdot f_{m})^{2}\;,
\end{align}
%\begin{equation}\label{eq.Ekin}
%E_{kin}=\dfrac{1}{2}\;J\;(2\cdot\pi\cdot f_{m})^{2},
%\end{equation}
where $J$ is the moment of inertia and $f_{m}$ is the rated rotational frequency of the machine. The inertia constant $H$ of a generator is defined as the ratio between the stored kinetic energy $E_{kin}$ and its rated power $S_{r}$~\cite{uriarte15}. $H$  determines the time interval during which an electrical generator can supply its rated power only by using the kinetic energy stored in its rotating masses~\cite{tielens16}:
\begin{align}
H=\dfrac{E_{kin}}{S_{r}}=\dfrac{J\;(2\cdot\pi\cdot f_{m})^{2}}{2\cdot S_{r}}\;.
\end{align}
%\begin{equation}
%H=\dfrac{E_{kin}}{S_{r}}=\dfrac{J\;(2\cdot\pi\cdot f_{m})^{2}}{2\cdot S_{r}}.
%\end{equation}

Depending on the type of conventional units (i.e., steam, combined cycle, hydroelectric, etc.), typical inertia constants are in the range of 2--10~s, as indicated in Table~\ref{tab.H}.
\begin{table}[!tb]
  \processtable{$H$ according to generation type, rated power and reference\label{tab.H}}
  {\begin{tabular*}{20pc}{@{\extracolsep{\fill}}lllll@{}}\toprule
     {\bf{Type of power plant}}    & {\bf{Rated power (MW)}} & {\bf{$H$ (s)}} & {\bf{Ref.}} & {\bf{Year}} \\
     %power plant & (MW)        &         &     &\\
     \midrule
     Thermal (2 poles) & Not indicated & 2.5-6 & \cite{kundur94} & 1994\\  
     Thermal (4 poles) & Not indicated & 4-10 & \cite{kundur94} & 1994\\ 
     Thermal & 10  & 4 &\cite{de07} & 2007\\  
     Thermal & 500-1500 & 2.3-2 & \cite{anderson08} & 2008\\ 
     Thermal & 1000  & 4-5 & \cite{dabur11} & 2011\\ 
     Thermal & Not indicated & 4-5 & \cite{kumal12} & 2012\\ 
     Thermal (steam) & 130 & 4 & \cite{tielens12} & 2012\\ 
     Thermal (steam) & 60 & 3.3 & \cite{tielens12} & 2012\\ 
     Thermal (combined cycle) & 115  & 4.3 & \cite{tielens12} & 2012\\ 
     Thermal (gas) & 90-120  & 5 & \cite{tielens12} & 2012\\ 
     Thermal (nuclear) & 100-1400  & 4 & \cite{tielens16} & 2016\\ 
     Thermal (fossil) & 0-1000  & 5-3 & \cite{tielens16} & 2016\\
     \midrule
     Hydroelectric & Not indicated & 2-4 & \cite{kundur94} & 1994\\
     Hydroelectric $n<$200 rpm& Not indicated & 2-3 &\cite{grainger94} & 1994\\ 
     Hydroelectric $n>$200 rpm & Not indicated & 2-4& \cite{grainger94} & 1994\\  
     Hydroelectric 450<n<514 rpm & 10-65  & 2-4.3& \cite{anderson08} & 2008\\ 
     Hydroelectric 200<n<400 rpm & 10-75 & 2-4& \cite{anderson08} & 2008\\ 
     Hydroelectric 138<n<180 rpm & 10-90  & 2-3.3& \cite{anderson08} & 2008\\ 
     Hydroelectric 80<n<120 rpm & 10-85  & 1.75-3& \cite{anderson08} & 2008\\ 
     Hydroelectric & Not indicated & 4.75 & \cite{eremia13} & 2013\\ 					
     \botrule
   \end{tabular*}}{}
\end{table}

\subsection{Swing equation of a power system. Equivalent rotational system inertia} \label{sec.swing_equation}
Power systems include several synchronous generators. Thus, it is possible to estimate the equivalent rotational system inertia ($H_{eq}$) by using~\cite{spahic16}:
\begin{align}
H_{eq}=\dfrac{\displaystyle\sum_{i=1}^{CP} H_{i}\cdot S_{B,i}}{S_{B}}\;,
\label{eq.Heq}
\end{align}
%\begin{equation}
%H_{R,eq}=\dfrac{\displaystyle\sum_{i=1}^{CP} H_{i}\cdot S_{B,i}}{S_{B}}  ,
%\label{eq.Heq}
%\end{equation}
$H_{i}$ refers to the inertia constant of power plant $i$, $S_{B,i}$ is the rated power of power plant $i$, $S_{B}$ is the rated power of the power system and $CP$ is the total number of conventional plants.

The swing equation of a power system is used to analyse transient stability problems, as well as frequency control design and regulation~\cite{raisz18}. Moreover, it relates frequency excursions with the power imbalance~\cite{tofis17}:
\begin{align}
\label{eq.swing}
\dfrac{d\Delta f}{dt} = \dfrac{1}{2\;H_{eq}} \left( \Delta P_{m}-\Delta P_{e}\right)\;,
\end{align}  
%\begin{equation}\label{eq.swing}
%\dfrac{d\Delta f}{dt} = \dfrac{1}{2\;H_{eq}} \left( \Delta P_{m}-\Delta P_{e}\right) \;
%\end{equation}
where $\Delta f$ is the deviation of the grid frequency, $H_{eq}$ is the equivalent inertia constant for the power system determined by~\eqref{eq.Heq}, $\Delta P_{m}$ is the mechanical power change supplied by generator and $\Delta P_{e}$ is the electrical power demand variation. 

Some electrical loads are frequency dependent (such as rotating machines). Consequently, $\Delta P_{e}$ is expressed as~\cite{suh17}:
\begin{align}
\label{eq.Pe}
\Delta P_{e} = \Delta P_{L}+D\cdot\Delta f\;,
\end{align}
%\begin{equation}\label{eq.Pe}
%\Delta P_{e} = \Delta P_{L}+D\cdot\Delta f\;,
%\end{equation}
being $\Delta P_{L}$ the power change of frequency independent loads and $D$ the damping factor (load-frequency response constant). Combining~\eqref{eq.swing} and~\eqref{eq.Pe}, the swing equation of a power system is obtained~\cite{yazdi19}.
\begin{align}
\label{eq.swing2}
\dfrac{d\Delta f}{dt} = \dfrac{1}{2\;H_{eq}} \left( \Delta P_{m}-\Delta P_{L}-D\;\Delta f\right) \;.
\end{align}
%\begin{equation}\label{eq.swing2}
%\dfrac{d\Delta f}{dt} = \dfrac{1}{2\;H_{eq}} \left( \Delta P_{m}-\Delta P_{L}-D\;\Delta f\right) \;.
%\end{equation}

\subsection{Future definition of inertia constant of a power system}\label{sec.current_definition}
By considering policies to promote the integration of renewables, RES
have replaced conventional power plants and, subsequently,
synchronous generators~\cite{li17design}. Among the different renewable sources available, PV and wind (especially doubly fed induction generators, DFIG~\cite{ochoa17}) are the two most promising resources for generating electrical energy~\cite{shah15}. Both wind and PV power plants are controlled by power converters according to the maximum power point tracking (MPPT) control~\cite{muyeen10,mohamed12}. This technique prevents both sources to directly contribute to the inertia of the system~\cite{zhao16,hosseinipour17,tielens17phd}, which is considered as one of the main drawbacks to integrate large amounts of RES into the grid~\cite{du18}. In fact, modern wind turbines have rotational inertia constants comparable to those of conventional generators, provided by their blades, drive train and electrical generator. However, this inertia is hidden from the power system point of view due to the converter~\cite{yingcheng11}. Moreover, ROCOF depends on the available
inertia~\cite{ulbig15}. As a result, larger frequency deviations are achieved after an imbalance between supply-side and
demand when RES replace conventional units without providing frequency response~\cite{nedd17}.

Therefore, it is necessary that RES become an active agent in grid frequency regulation~\cite{you17}. Actually, several TSOs are requiring that RES contribute to ancillary
services as well~\cite{aho12}, especially wind power
plants~\cite{kayikcci09}. Toulabi \textit{et al}. affirm that the
participation of wind turbines in frequency control is
necessary~\cite{toulabi17}. Under these requirements, different solutions providing
inertia and frequency control from RES have been under study during
the last decades. These technologies are usually known as `virtual inertia techniques'~\cite{tamrakar17} and are explained in~\cite{sun10,tamrakar17,attya18,wang18,ziping18}. 
%\begin{figure*}[!t]
%	\centering
%	\includegraphics[width=0.475\textwidth]{figures-def/esquema.pdf}
%	\caption{Inertia and frequency control techniques for RES}
%	\label{fig.esquema}
%\end{figure*}

If RES providing frequency response were considered, the equivalent inertia of the power system
would have two different components: $(i)$ synchronous rotational inertia due to
conventional generators $H_{R,eq}$, calculated with eq.~\eqref{eq.Heq}, and $(ii)$ virtual inertia
corresponding to RES, $H_{V,eq}$, as indicated in
eq.~\eqref{eq.hagg2}~\cite{morren06phd,tielens17}. In this way,
$H_{V,j}$ refers to the emulated inertia constant of power plant $j$, $S_{B,j}$ is the rated power of power plant $j$
and $VG$ is the total number of virtual generators included in the
power system under consideration. The rest of the parameters are the same as~\eqref{eq.Heq}.
\begin{align}
H_{eq}=\dfrac{\overbrace{\displaystyle\sum_{i=1}^{CP}H_{i}\cdot S_{B,i}}^{H_{R,eq}}+\overbrace{\displaystyle\sum_{j=1}^{VG}H_{V,j}\cdot S_{B,j}}^{H_{V,eq}}}{S_{B}}
\label{eq.hagg2}
\end{align}
%\begin{equation}
%H_{eq}=\dfrac{\overbrace{\displaystyle\sum_{i=1}^{CP}H_{i}\cdot S_{B,i}}^{H_{R,eq}}+\overbrace{\displaystyle\sum_{j=1}^{VG}H_{V,j}\cdot S_{B,j}}^{H_{V,eq}}}{S_{B}}
%\label{eq.hagg2}
%\end{equation}
However, the values of $H_{V,j}$ are not normally known and can be time dependent. Thus, it is difficult to apply eq.~\eqref{eq.hagg2}.

\section{Inertia estimation strategies. Methodology}\label{sec.inertia_estimation}
Different inertia estimation strategies have been proposed during the last decades~\cite{inoue97,chassin05,wall12,wall14,zografos17,tuttelberg18,zografos18}. Damping factor is neglected in most approaches as its effects are small on the firsts moments of the imbalance $\Delta P$.

Inoue \textit{et al.} propose a procedure for estimating the inertia constant of a power system using transients of the frequency measured at an imbalance~\cite{inoue97}. At the onset of an imbalance $(t=0^{+})$, the frequency deviation is $\Delta f=0$. Assuming that the imbalance $\Delta P=\Delta P_{m}-\Delta P_{L}$ is known, and by estimating the ROCOF ($df/dt$) at $t=0^{+}$, the inertia constant can be calculated with
\begin{align}
\label{eq.inoue}
H_{eq}=\dfrac{-\Delta P}{\left.\dfrac{2\;d(\Delta f/f_{0})}{dt}\right| _{t=0^{+}}}\;.
\end{align}
%\begin{equation}\label{eq.inoue}
%H_{eq}=\dfrac{-\Delta P}{\left.\dfrac{2\;d(\Delta f/f_{0})}{dt}\right| _{t=0^{+}}}\;.
%\end{equation}
To calculate the ROCOF, a $5^{th}$ degree polynomial approximation of $\Delta f/f_{0}$ with respect to time is fitted. The time interval is about 15 to 20~s after the imbalance
\begin{align}
\Delta f/f_{0}=A_{5}\cdot t^{5}+A_{4}\cdot t^{4}+A_{3}\cdot t^{3}+A_{2}\cdot t^{2}+A_{1}\cdot t\;,
\end{align}
%\begin{equation}
%\Delta f/f_{0}=A_{5}\cdot t^{5}+A_{4}\cdot t^{4}+A_{3}\cdot t^{3}+A_{2}\cdot t^{2}+A_{1}\cdot t\;,
%\end{equation}
where $t$ is the time. By estimating the coefficients $A_{1}$ to $A_{5}$, the equivalent inertia constant $H_{eq}$ is obtained by using eq.~\eqref{eq.inoue97}, as $A_{1}$ is approximately equal to the ROCOF at $t_{0}^{+}$
\begin{align}
A_{1}=f'(t=0^{+})\approx \left.\dfrac{\Delta f/f_{0}}{dt}\right|_{t=0}
\end{align}
%\begin{equation}
%A_{1}=f'(t=0^{+})\approx \left.\dfrac{\Delta f/f_{0}}{dt}\right|_{t=0}
%\end{equation} 
\begin{align}
\label{eq.inoue97}
H_{eq}=\dfrac{-\Delta P}{2\cdot A_{1}}\;.
\end{align}
%\begin{equation}\label{eq.inoue97}
%H_{eq}=\dfrac{-\Delta P}{2\cdot A_{1}}\;.
%\end{equation}

Chassin \textit{et al}~\cite{chassin05} frequency and power values from the Western Electricity Coordination Council were collected. In this case, ROCOF is estimated by removing noise from the frequency data recorded and applying the first derivative. The equation to estimate $H_{eq}$ is as below
\begin{align}
\label{eq.chassin}
H_{eq}=\dfrac{-\Delta P}{2\,\dfrac{df}{dt}}\;.
\end{align}
%\begin{equation}\label{eq.chassin}
%H_{eq}=\dfrac{-\Delta P}{2\,\dfrac{df}{dt}}\;.
%\end{equation}

\begin{figure}[!tb]
\centering
\includegraphics[width=0.35\textwidth]{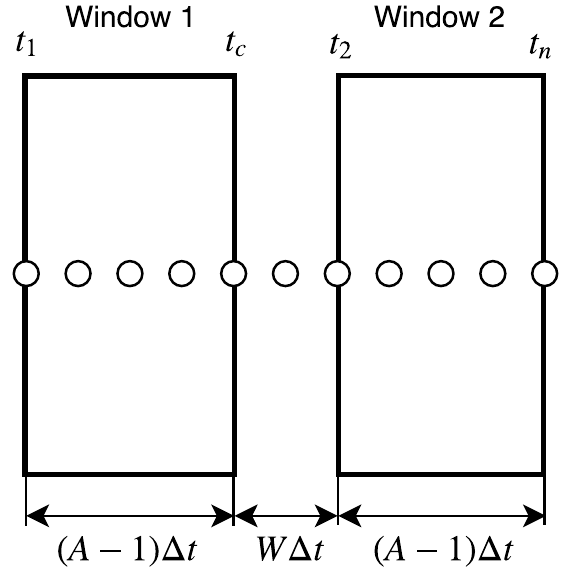}
\caption{Sample of windows. In this case, $A=5$ and $W=2$\label{fig.windows}}
\end{figure}

Wall \textit{et al.} present a robust estimation method for the inertia available in the system~\cite{wall12,wall14}. It uses as input data the active power $P$ and the derivative of frequency $df(t)/dt$, measured from a single location. The  proposed algorithm consists of a set of four filters (two for the total active power --$P_{1}$ and $P_{2}$-- and two for the ROCOF --$R_{1}$ and $R_{2}$--) applied as sliding windows, see Figure~\ref{fig.windows}. Windows have a width of $A$ data points and they are separated by a width $W$. 

$H_{eq}$ is estimated by the following expression:
\begin{align}
\label{eq.wall}
H_{eq}=\dfrac{1}{2}\dfrac{P_{1}-P_{2}}{R_{2}-R_{1}}\;,
\end{align}
where $P_{1}$, $P_{2}$, $R_{1}$ and $R_{2}$ are calculated with~\eqref{eq.wall2}:
\begin{align}
\begin{split}
P_{1}(t_{n})=\dfrac{1}{A}\sum_{t=t_{1}}^{t_{c}}P(t)\;,\\
P_{2}(t_{n})=\dfrac{1}{A}\sum_{t=t_{2}}^{t_{n}}P(t)\;,\\
R_{1}(t_{n})=\dfrac{1}{A}\sum_{t=t_{1}}^{t_{c}}\dfrac{df(t)}{dt}\;,\\
R_{2}(t_{n})=\dfrac{1}{A}\sum_{t=t_{2}}^{t_{n}}\dfrac{df(t)}{dt}\;.
\end{split}
\label{eq.wall2}
\end{align}
The result of applying eq.~\eqref{eq.wall} is only $H_{eq}$ during the time in which the power imbalance has occurred ($t_{dist}$)~\cite{wall14}.

Zografos and Ghandhari~\cite{zografos17} consider an aggregated load model to represent the behaviour of the average system load. The load power change is expressed by
\begin{align}
\label{eq.zografos17_1}
\Delta P_{L}(t)=P_{prod}\cdot \left( V_{s}(t)-1\right) \;
\end{align}
where $P_{prod}$ is the total power production before the disturbance and $V_{s}(t)$ is the system's overall voltage profile, approximated by the voltage of the generator buses according to
\begin{align}
\label{eq.zografos17_2}
V_{s}(t)=\dfrac{\displaystyle\sum_{i=1}^{n}\left( \dfrac{V_{G,i}(t)}{V_{G0,i}}\right)}{n} \;,
\end{align}
being $V_{G,i}(t)$ the voltage at the bus of generator $i$ at time $t$, $V_{G0,i}$ the voltage before the disturbance at the bus of generator $i$ and $n$ the number of connected generators. By combining~\eqref{eq.swing2} and~\eqref{eq.zografos17_1}, the inertia constant of the system is calculated from~\eqref{eq.zografos17}, where $\Delta P_{dist}$ is the size of the disturbance at the moment of the disturbance
\begin{align}
\label{eq.zografos17}
H_{est}=\dfrac{\Delta P(t)}{2\cdot \dfrac{df}{dt}}=\dfrac{\Delta P_{L}(t)+\Delta P_{dist}}{2\cdot \dfrac{df}{dt}}\;.
\end{align}

Tuttelberg \textit{et al.} ~\cite{tuttelberg18} simplify the dynamic response to a reduced order system with the generic form of~\eqref{eq.tuttelberg}
\begin{align}
\label{eq.tuttelberg}
H(s)=\dfrac{b_{n-1}s^{n-1}+b_{n-2}s^{n-2}+...+b_{0}}{a_{n}s^{n}+a_{n-1}s^{n-1}+...+a_{0}}\;.
\end{align}
The inertia of a power system $H_{eq}$ can be determined by the value of its unit impulse response at $t=0$. For a transfer function like the one presented in~\eqref{eq.tuttelberg}, the first value of the impulse response can be evaluated in Matlab with: $(i)$~the {\ttfamily{impulse}} function, $(ii)$~the gain value of the zero-pole model from {\ttfamily{tf2zpk}} or $(iii)$~as the ratio of $a_{n}$ to $-b_{n-1}$.

Zografos \textit{et al.}~\cite{zografos18} introduce two approaches to express the power change due to the frequency and voltage dynamics ($R$ and $V$ approaches, respectively)
\begin{align}
\label{eq.zografos18}
\Delta P(t)=h_{1}(f(t))+h_{2}(V(t))-\Delta P_{dist}\;,
\end{align}
where $P_{dist}$ is the size of the disturbance, and $h_{1}(f(t))$ and $h_{2}(V(t))$ deal with the power change due to the frequency and the voltage dynamics, respectively.

In the $R$ approach, it is considered that $\Delta P(t)=h_{1}(f(t))-\Delta P_{dist}$. To obtain $h_{1}(f(t))$, the governor's behavior is analysed. $h_{1}(f(t))$ relates the mechanical power change and the frequency deviation. It is considered that
\begin{align}
\label{eq.zografos18_2}
\Delta P_{m}(t)=-R(t)\cdot\Delta f(t)\;,
\end{align}
being $\Delta P_{m}$ the mechanical power change and $R(t)$ an unknown time varying function that accommodates the dynamic response of the system related to $\Delta f(t)$. Then eq.~\eqref{eq.swing2} is converted into
\begin{align}
\label{eq.zografos18_3}
2\cdot H_{eq}\dfrac{df}{dt}=h_{1}(f(t))-\Delta P_{dist}=R(t)\cdot\Delta f(t)-\Delta P_{dist}\;
\end{align}
where $H_{eq}$ is the estimated inertia constant to be found. However, as previously said, $R(t)$ is also unknown. To compute $R(t)$, a specific selected time $t_{sr}$ is considered. $t_{sr}$ is recommended to be the first local extreme of the ROCOF curve after the moment of the disturbance. Moreover, eq.~\eqref{eq.zografos18_3} is considered for $N$ discrete points equally distributed around $t_{sr}$. $R(t)$ can thus be approximated by the average of the values of $R(t)$ of the $N$ neighbouring points to $t_{sr}$. Therefore, a system with $N+1$ linear equations and $N+1$ unknowns is obtained~\eqref{eq.zografos18_4}. By solving it, $R(t_{sr})$ is obtained
\begin{align}
\begin{gathered}
\label{eq.zografos18_4}
%\begin{split}
2\cdot H_{eq}\dfrac{df(t_{sr}+i)}{dt} =  R(t_{sr}+i)\cdot\Delta f(t_{sr}+i)-\Delta P_{dist} \\
R(t_{sr})=  \dfrac{\displaystyle\sum_{i=-N/2}^{N/2}R(t_{sr}+i)}{N}\\
\forall i \in \Z:  -N/2\leq i\leq N/2 : i\neq0  
%\end{split}
\end{gathered}
\end{align}

In the $V$ approach, it is considered that $\Delta P(t)=h_{2}(V(t))-\Delta P_{dist}$. To obtain $h_{2}(V(t))$, the load power change due to voltage dependency is analysed
\begin{align}
\label{eq.zografos18_5}
\Delta P_{LV}(t)=P_{prod}(k_{z}(V_{s}(t)))^{2}+k_{i}(V_{s}(t)+k_{p})-P_{prod}\;,
\end{align} 
where $P_{prod}$ is the total power production before the disturbance, $k_{z}$, $k_{i}$ and $k_{p}$ define the fraction of each component, and $V_{s}(t)$ is the loads' aggregated voltage profile, calculated with~\eqref{eq.zografos17_2}. Then
\begin{align}
\label{eq.zografos18_6}
2\cdot H_{eq}\dfrac{df}{dt}=h_{2}(V(t))-\Delta P_{dist}=-\Delta P_{LV}(t)-\Delta P_{dist}\;.
\end{align}
The application range $t_{sv}$ of this strategy should be selected before 500~ms, and as soon as possible after the disturbance to avoid the governor frequency response.

The estimated equivalent inertia is calculated with~\eqref{eq.zografos18_7}, where $t_{s}$ is recommended to be the $t_{sr}$ estimated with~\eqref{eq.zografos18_4}
\begin{align}
\label{eq.zografos18_7}
H_{eq}=\dfrac{ R(t_{s})\Delta f(t_{s})-\Delta P_{LV}(t_{s})-\Delta P_{dist}}{2\dfrac{df(t_{s})}{dt}}
\end{align}

Finally, Table~\ref{tab.methodologies} summarizes the different
inertia estimation methodologies discussed in this work. As can be
seen, most of them are based on the power imbalance and ROCOF, in line
with the swing equation and the frequency control of conventional
generation units.

\begin{table}[!tbp]
  \processtable{Summary of inertia estimation methodologies\label{tab.methodologies}}
  {\begin{tabular*}{20pc}{@{\extracolsep{\fill}}lll@{}}\toprule
     {\bf{Ref.}}  & {\bf{Methodology based on }} & {\bf{Year}} \\
     %      & based on    &\\
     \midrule
     \cite{inoue97} & Power imbalance and ROCOF & 1997  \\
     \cite{chassin05} & Power imbalance and ROCOF & 2005  \\
     \cite{wall12} & Total power supplied and ROCOF & 2012  \\
     \cite{wall14} & Total power supplied and ROCOF & 2014\\
     \cite{zografos17} & Power imbalance and ROCOF & 2017\\
     \cite{tuttelberg18} & Impulse function & 2018\\
     \cite{zografos18} & Power imbalance and ROCOF & 2018\\		
     \botrule
   \end{tabular*}}{}
\end{table}

\section{System identification}\label{sec.system_identification}

\subsection{Power system modelling} \label{sec.power_system_modeling}

\begin{figure}[!tb]
  \centering
  \includegraphics[width=0.5\textwidth]{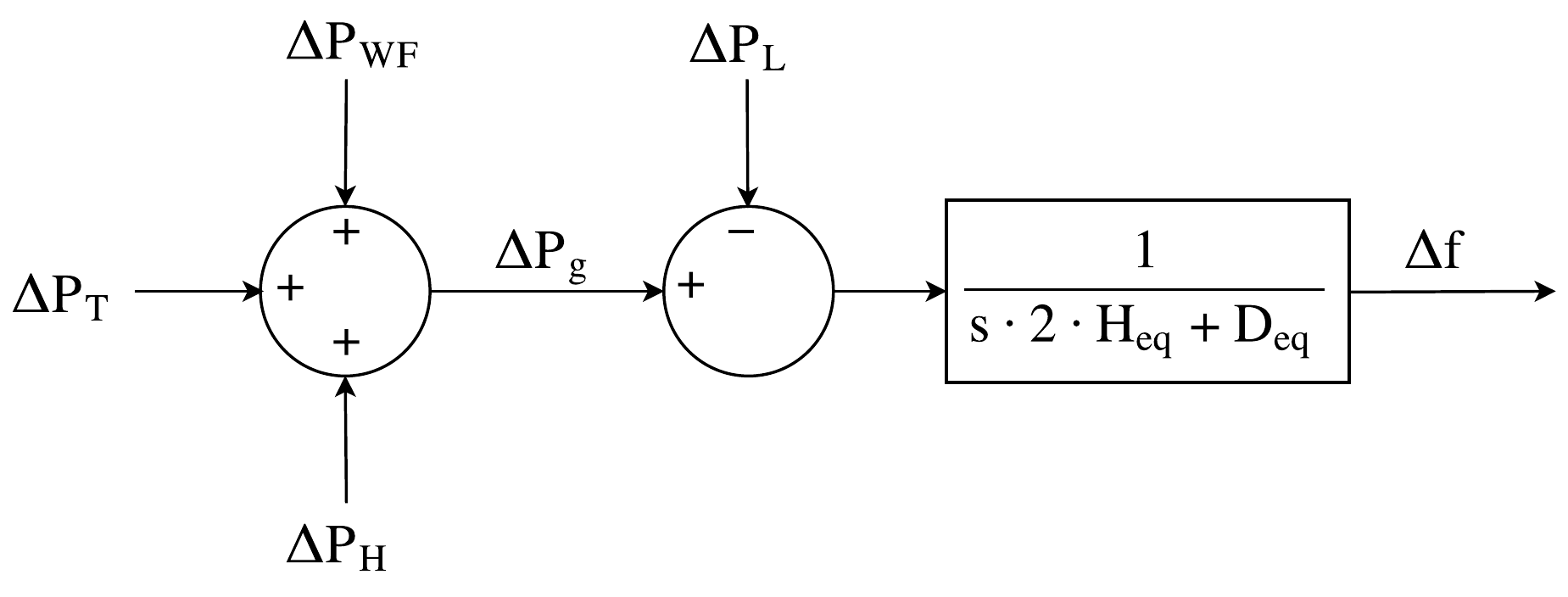}
  \caption{\color{black}Simplified diagram of the electrical power system used for simulations~\cite{benavente19}\label{fig.model}}
\end{figure}

{\color{black}From the supply-side, the power system considered for simulation
purposes involve conventional generating units (thermal and
hydro-power plants) and wind power plants. A simplified diagram of the
power system can be seen in Figure~\ref{fig.model}, being the variation of the generated power $\Delta P_{g} = \Delta P_{WF} + \Delta P_{T} + \Delta P_{H}$, and $\Delta P_{L}$ the power imbalance. A base power of 1350~MW is
assumed, corresponding to the capacity of the power system. It is considered that the active power of loads is
independent on voltage, and as a consequence, the term $\Delta P_{L}(t)$
of eq.~\eqref{eq.zografos17}~\cite{zografos17} is not considered, and
the $V$ approach of Zografos~\textit{et al.}~\cite{zografos18} is not
taken into account. The equivalent damping factor of loads is
$D_{eq}=1$~pu$_{MW}$/pu$_{Hz}$~\cite{kundur94}. Simulations have been
carried out in Matlab/Simulink.}

\begin{figure}[!tb]
  \centering
  \subfloat[]{\includegraphics[width=.95\linewidth]{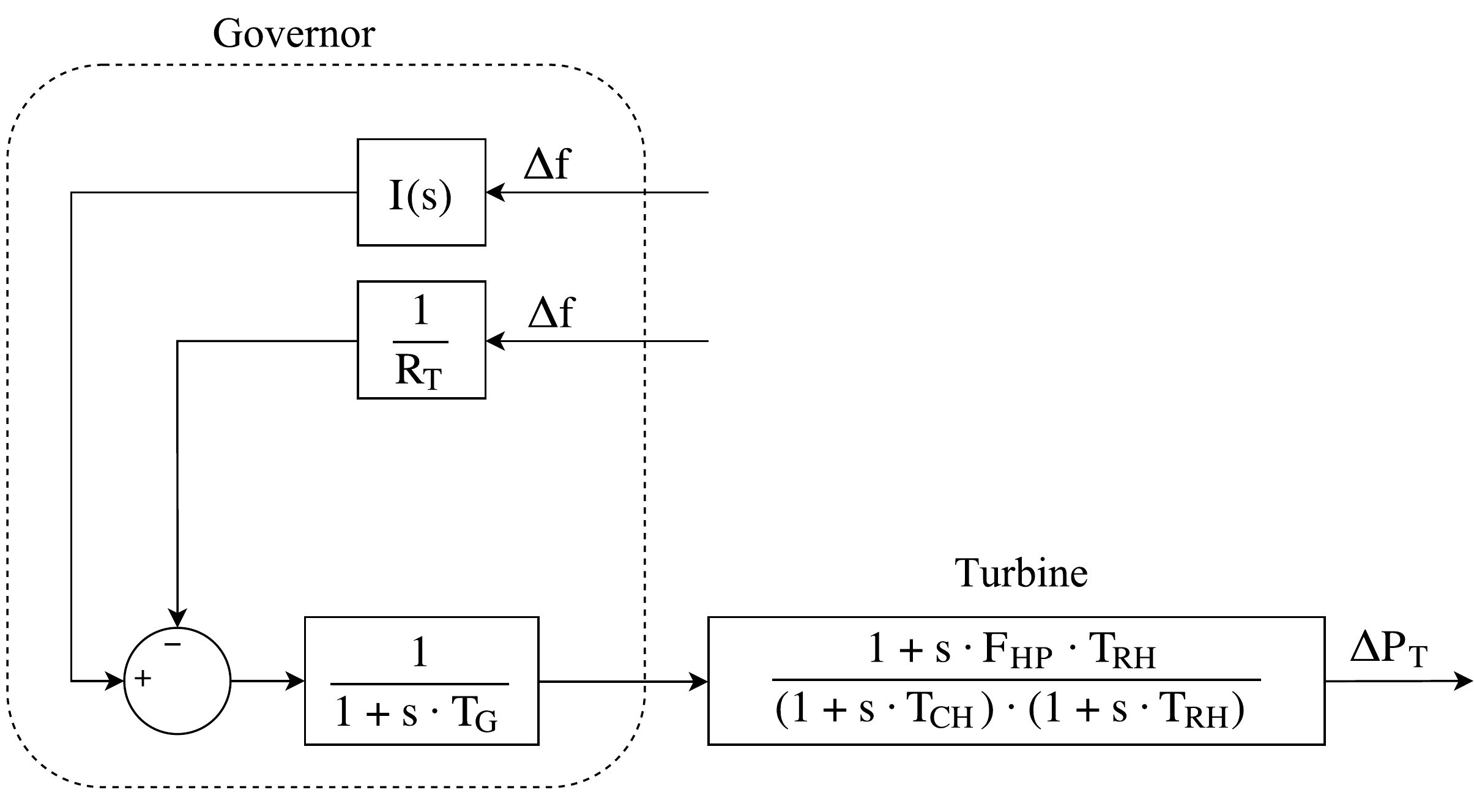}\label{fig.thermal}}\\
  \subfloat[]{\includegraphics[width=.95\linewidth]{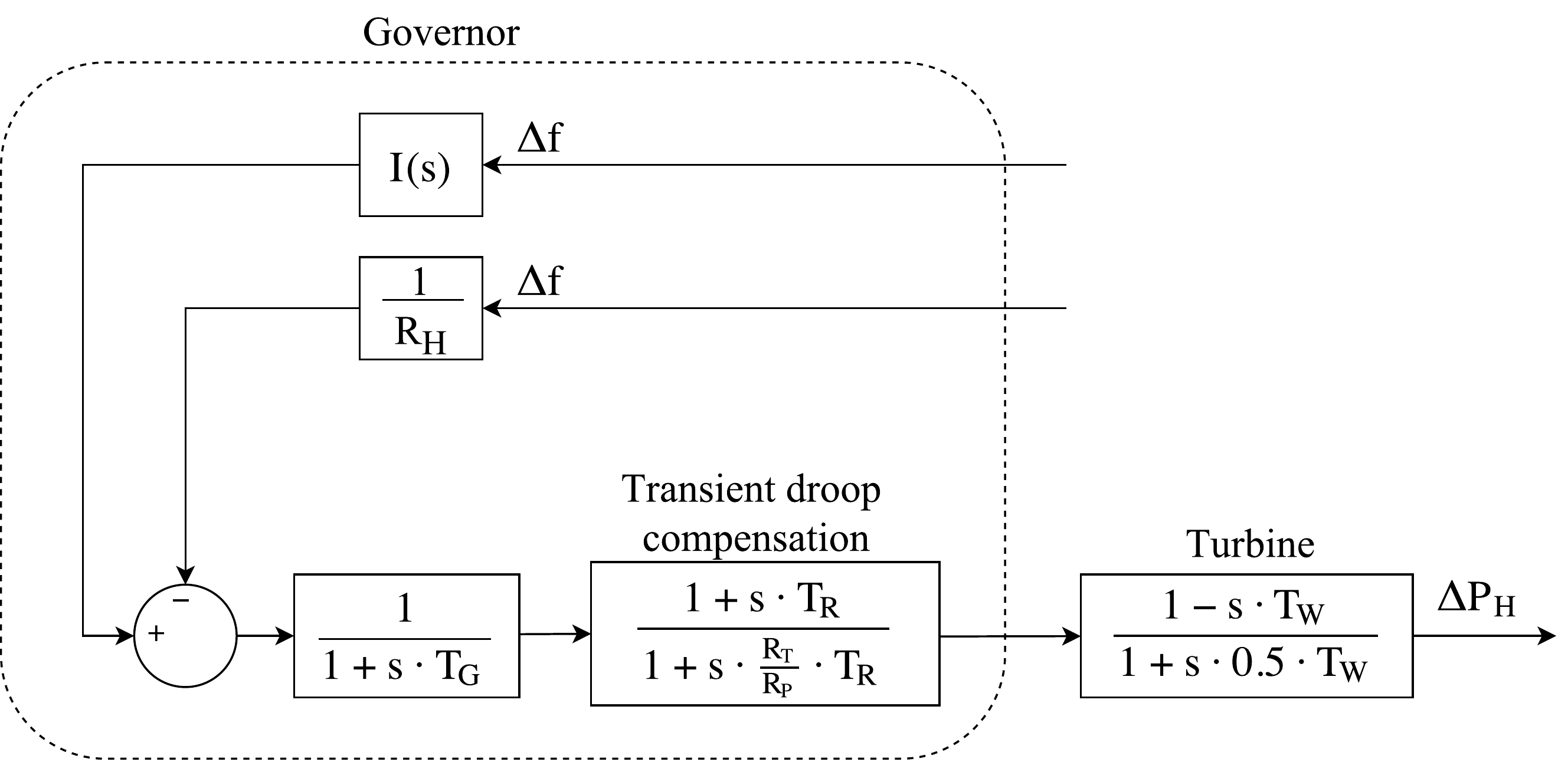}\label{fig.hydro}}
  \caption{\color{black}Thermal and hydro-power plant models~\cite{kundur94}}
  \figfooter{\color{black}(a)}{\color{black}Reheat thermal power plant}
  \figfooter{\color{black}(b)}{\color{black}Hydro-power}
  \label{fig.conventional}
\end{figure}

{\color{black}Conventional units are modelled according to the simplified
governor-based models widely used and proposed in~\cite{kundur94}, see Figure~\ref{fig.conventional}. The
inertia constant for these power plants are $H_{thermal}=5$~s and $H_{hydro}=3.3$~s. Wind power plants are
modelled according to an equivalent wind turbine, with the mechanical
single-mass and turbine control models presented
in~\cite{miller03,ullah08,clark10}. The frequency controller is included in the
wind turbine model as can be seen in Figure~\ref{fig.aero_control}. Parameters of both conventional and wind power plants are summarized in the Appendix.}

\begin{figure*}[!tb]
  \centering
  \includegraphics[width=0.72\textwidth]{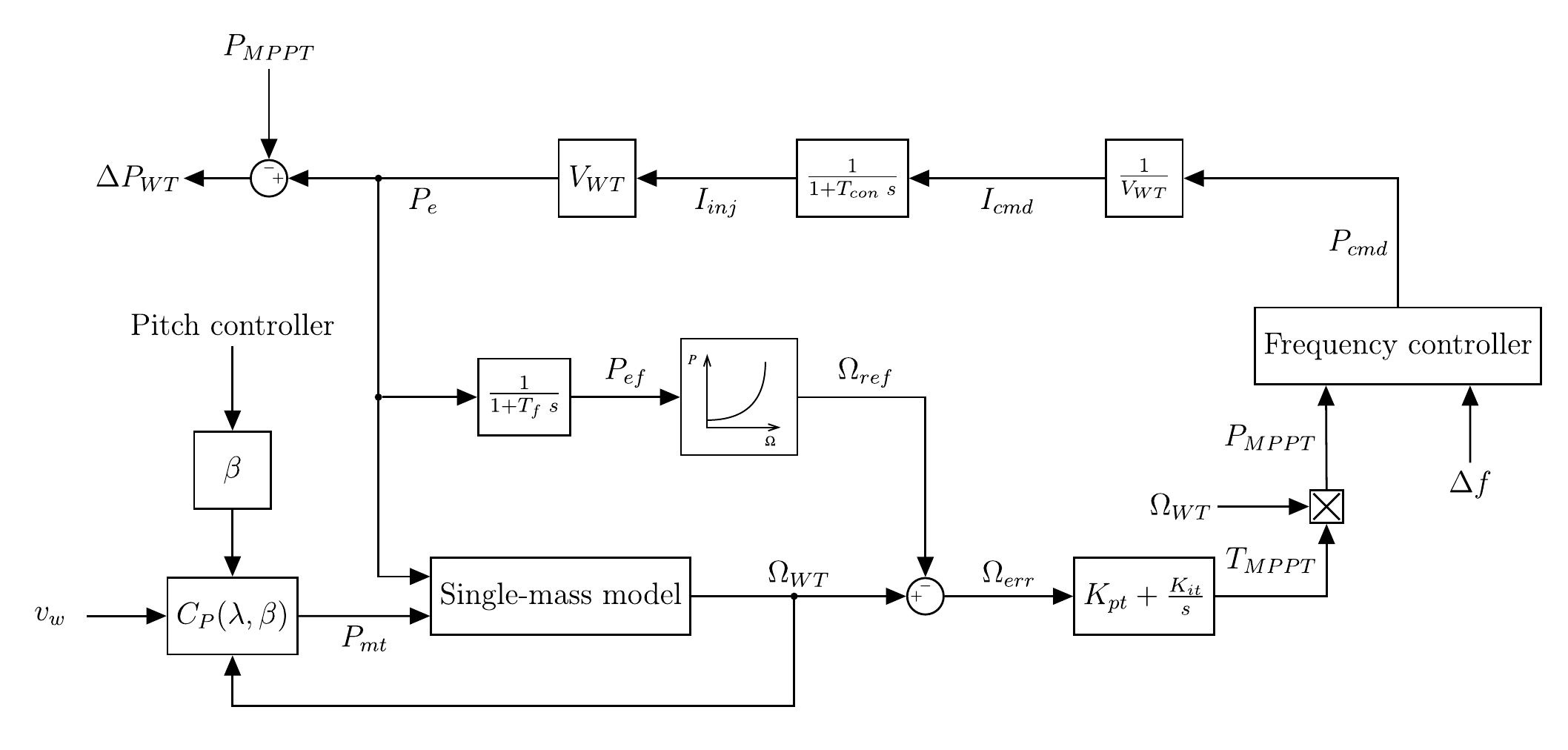}
  \caption{Variable speed wind turbine model with frequency controller\label{fig.aero_control}}
\end{figure*}

\subsection{Frequency control strategies} \label{sec.frequency_control} 

Under power imbalance
conditions, the governor control mechanisms of conventional units
modify their active power supply to recover system power balance and,
thus, remove the frequency deviation~\cite{diaz14}. Grid frequency deviation
$\Delta f$ is subsequently used as an input signal for primary and secondary
frequency controls~\cite{alomoush10}. Primary frequency control is
performed locally at the generator, being the active power
increment/decrement proportional to $\Delta f$ through the speed
regulation parameter $R$~\cite{dai12}. Secondary frequency control
involves an integral controller that modifies the turbine set-point of
each generation unit~\cite{simpson15}.

\begin{figure}[!tb]
  \centering
  \includegraphics[width=0.5\textwidth]{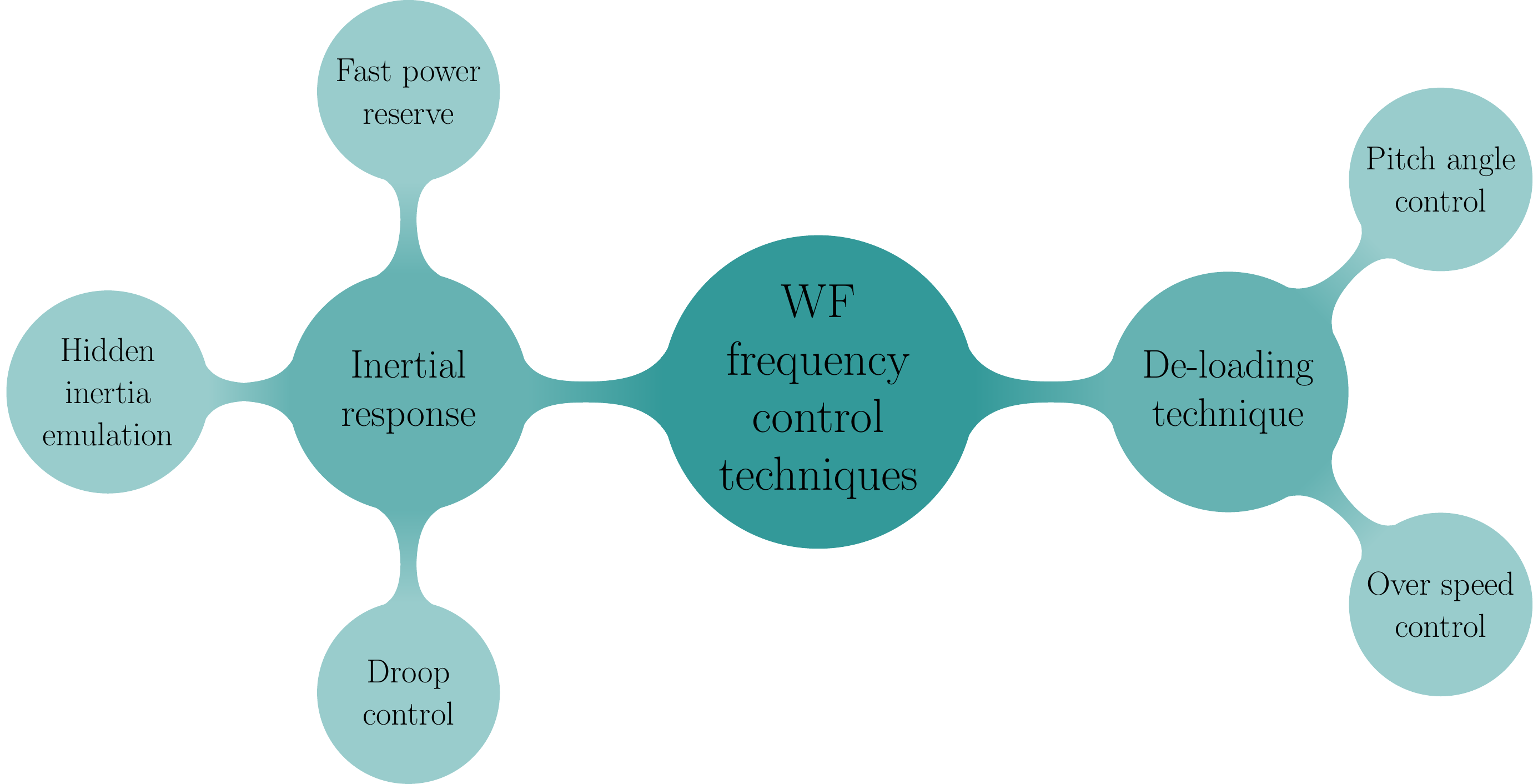}
  \caption{Frequency control techniques for wind power plants\label{fig.esquema}}
\end{figure}

{\color{black}Wind turbines can also include frequency control
  strategies. Different solutions have been proposed in the last
  decade. These strategies are usually classified as indicated in
  Figure~\ref{fig.esquema}~\cite{dreidy17}, excluding the use of
  energy storage systems. According to the specific literature,
  examples of these strategies are summarized in
  Table~\ref{tab.freq_wt}.}

Moreover, some approaches can be combined, in order to improve the
frequency deviation after the power
imbalance~\cite{margaris12,zhang12,ye15,abo16,hwang16,van16,liu19}. As
can be seen, an alternative classification can be then proposed:
$(i)$~not-including derivative frequency dependence and
$(ii)$~including derivative frequency dependence. An additional active
power $\Delta P$ is added to the pre-event power supplied by the wind
power plant $P_{0}$ in all the cases except de-loading technique. In
the fast power reserve, $\Delta P$ can be defined: $(i)$ as a
constant, $(ii)$ proportional to the rotational speed of the turbine
or $(iii)$ proportional to the frequency excursion, depending on the
reference. The hidden inertia emulation uses a proportional derivative
controller, being $K_{d}$ and $K_{p}$ the derivative and proportional
constants of the controller, respectively. With regard to the droop
control, $\Delta P$ is proportional to the frequency deviation
$\Delta f$ by the droop constant~$R$. As discussed in Section
\ref{sec.results}, this frequency controllers modify considerably the
estimated inertia values and addresses significant discrepancies among
methodologies.

\begin{table}[!tbp]
  \processtable{Wind turbines frequency control proposals\label{tab.freq_wt}}
  {\begin{tabular*}{20pc}{@{\extracolsep{\fill}}llll@{}}
     \toprule
     {\bf{Ref.}} & {\bf{Type of control}} & {\bf{Definition}} & {\bf{Year}} \\
     \midrule
     \cite{tarnowski09} & Fast power reserve & $P_{0}+\Delta P$, $\Delta P=cte$ & 2009\\
     \cite{keung09} & Fast power reserve & $P_{0}+\Delta P$, $\Delta P=cte$ & 2009\\
     \cite{chang11}& Fast power reserve & $P_{0}+\Delta P$, $\Delta P\propto\Omega$   & 2011 \\ 
     \cite{itani11}& Fast power reserve & $P_{0}+\Delta P$, $\Delta P=cte$ & 2011\\
     \cite{hansen14}& Fast power reserve & $P_{0}+\Delta P$, $\Delta P=cte$ & 2014 \\ 
     \cite{kang15}& Fast power reserve &$P_{0}+\Delta P$, $\Delta P=cte$ & 2015 \\
     \cite{hafiz15}& Fast power reserve & $P_{0}+\Delta P$, $\Delta P=cte$ & 2015\\
     \cite{kang16}& Fast power reserve	& $P_{0}+\Delta P$, $\Delta P\propto \Omega$& 2016 \\  
     % \cite{zhang17}& Fast power reserve & asdf  & 2017 \\ 
     \cite{fernandez18}& Fast power reserve& $P_{0}+\Delta P$, $\Delta P\propto \Delta f$& 2018 \\
     \midrule
     \cite{su12}& Hidden inertia emulation &$P_{0}+K_{d}\;df/dt+K_{p}\Delta f$ & 2012 \\
     \cite{zhang12} & Hidden inertia emulation &$P_{0}+K_{d}\;df/dt+K_{p}\Delta f$  & 2012\\
     \cite{zhang13} & Hidden inertia emulation & $P_{0}+K_{d}\;df/dt+K_{p}\Delta f$ & 2013\\
     \cite{you15}& Hidden inertia emulation &$P_{0}+K_{d}\;df/dt+K_{p}\Delta f$ & 2015 \\		
     \cite{hwang16}& Hidden inertia emulation &$P_{0}+K_{d}\;df/dt+K_{p}\Delta f$  & 2016 \\
     \midrule
     \cite{bonfiglio16} & Droop & $P_{0}+R\;\Delta f$ & 2016\\
     \cite{persson16} & Droop & $P_{0}+R\;\Delta f$ & 2016\\
     \cite{ye17} & Droop & $P_{0}+R\;\Delta f$ & 2017\\
     \cite{jahan19} & Droop & $P_{0}+R\;\Delta f$ & 2019 \\
     \midrule
     \cite{wilches16} & Pitch angle deloading & --- & 2016 \\ 
     \botrule
   \end{tabular*}}{}
\end{table}

\begin{figure}[!tbp]
  \centering
  \subfloat[]{\includegraphics[width=.95\linewidth]{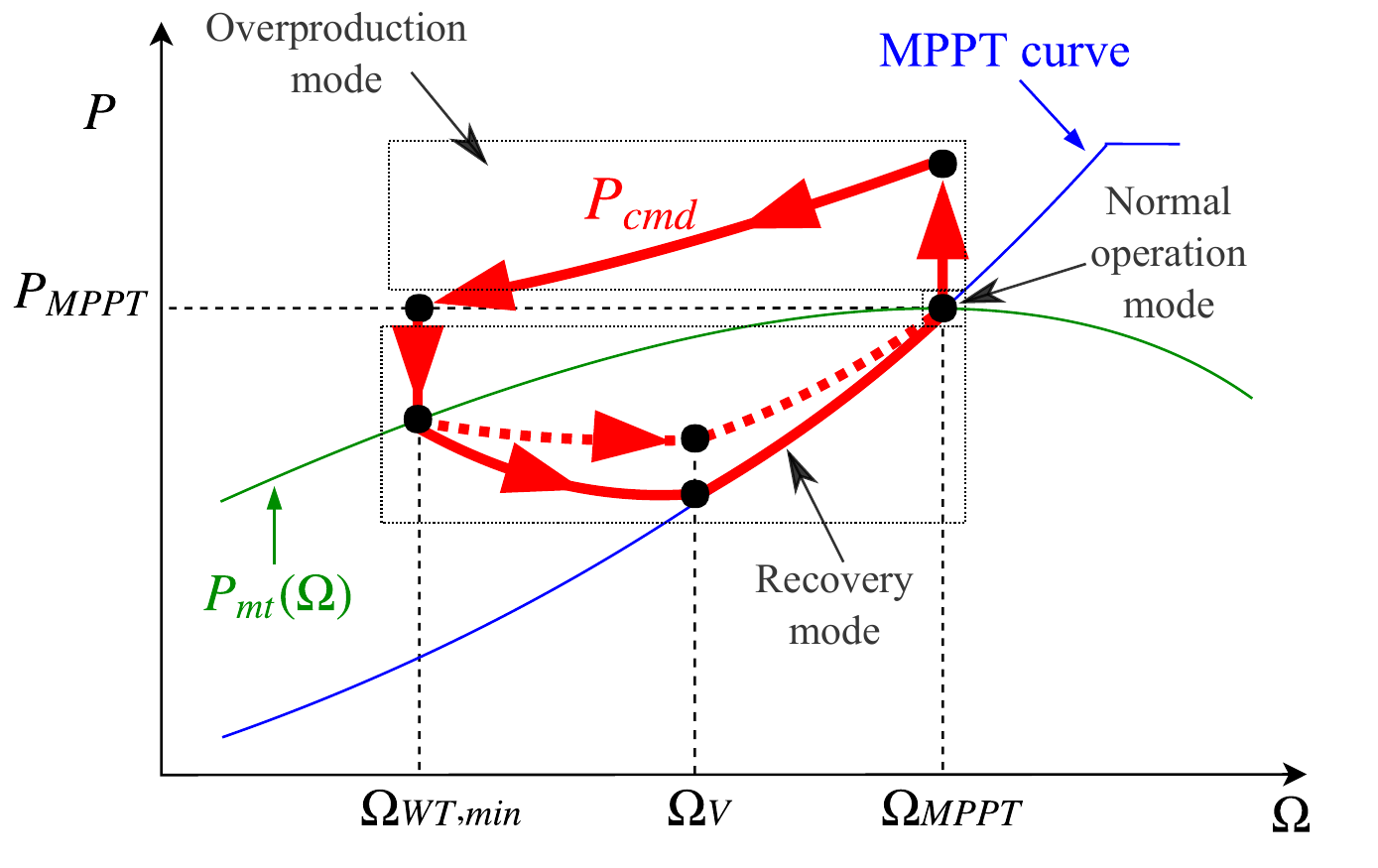}\label{fig.control}}\\
  \subfloat[]{\includegraphics[width=.95\linewidth]{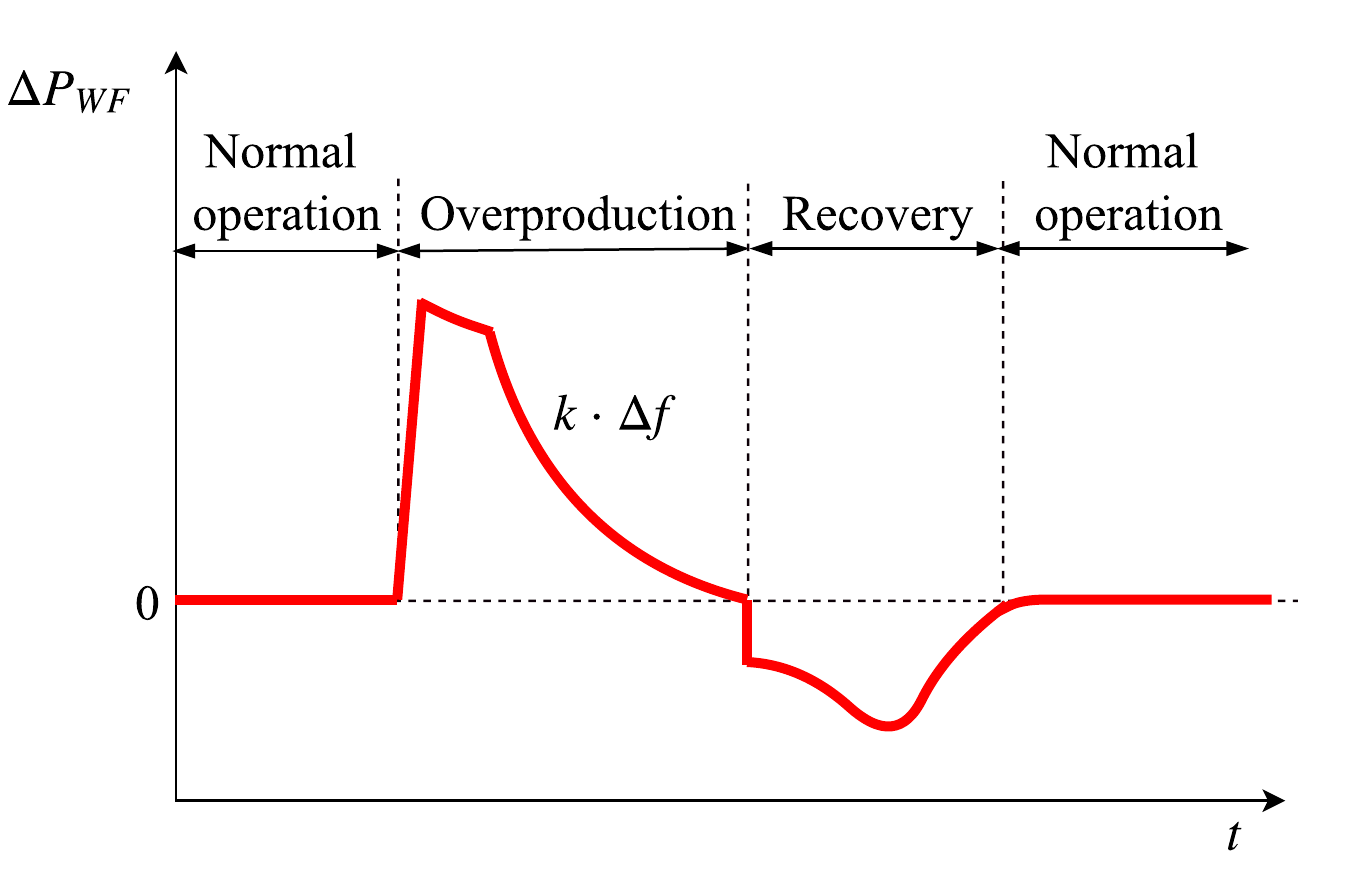}\label{fig.power}}
  \caption{Wind frequency control strategy and VSWTs active power variation \cite{fernandez18}.}
  \figfooter{\color{black}(a)}{{\color{black}Frequency control strategy}}
  \figfooter{(b)}{$\Delta P_{WF}$ with frequency control strategy}
  \label{fig.tfcf}
\end{figure}

{\color{black}The strategy for VSWTs implemented in this paper is
  based on the fast power reserve technique presented
  in~\cite{fernandez18} for for isolated power systems and assessed
  in~\cite{fernandez18fast} for multi-area power systems. As indicated
  in Figure~\ref{fig.tfcf}, under power imbalance conditions three
  operation modes are considered: $(i)$~normal operation mode,
  $(ii)$~overproduction mode and $(iii)$~recovery mode. Different
  commanded active power ($P_{cmd}$) values are determined aiming to
  restore the grid frequency. Figure~\ref{fig.control} depicts the
  trajectory of $P_{cmd}$ in a $\Omega_{WT}-P$ plot, indicating the
  three different operation modes. In Figure~\ref{fig.power}, the
  VSWTs active power variations ($\Delta P _{WF}$) submitted to an
  under-frequency excursion can be seen, being
  $\Delta P_{WF}=P_{cmd}-P_{MPPT}(\Omega_{MPPT})$. }

\begin{enumerate}[i. ]
	\item In the {\em{normal operation mode}}, the VSWTs operate at the maximum available active power for the current wind speed $P_{MPPT}(v_{w})$ and the available mechanical power ($P_{mt}(\Omega_{WT})$).
	\begin{align}
	P_{cmd}=P_{mt}(\Omega_{WT})=P_{MPPT}(v_{w}).
	\end{align}
	%	\begin{equation}
	%	P_{cmd}=P_{mt}(\Omega_{WT})=P_{MPPT}(v_{w}).
	%	\end{equation}
	When a generation-load mismatch occurs, the frequency controller strategy switches to the
	overproduction mode,\\
	\begin{align}
	|\Delta f|>\Delta f_{lim}\rightarrow \mathrm{Overproduction}.
	\end{align}
	%	\begin{equation}
	%	\Delta f<-\Delta f_{lim}\rightarrow \mathrm{Overproduction}.
	%	\end{equation}
	\item In the {\em{overproduction mode}}, the active power supplied by the
	VSWTs ($P_{cmd}$) is over the available mechanical power $P_{mt}(\Omega_{WT})$ curve. The additional active power
	$\Delta P_{OP}$ is provided by the kinetic energy stored in the
	rotational masses, and is proportional to $\Delta f$ to emulate primary frequency control of
	conventional generation
	units~\cite{margaris12}.
	\begin{align}
	P_{cmd}=P_{mt}(\Omega_{WT})+\Delta P_{OP}(\Delta f).
	\end{align}
	%	\begin{equation}
	%	P_{cmd}=P_{mt}(\Omega_{WT})+\Delta P_{OP}(\Delta f).
	%	\end{equation}
	Overproduction mode remains active until: either the
	rotational speed reaches a minimum allowed value $\Omega_{WT,min}$ or the commanded
	power $P_{cmd}$ is lower than the maximum available active power $P_{MPPT}(\Omega_{MPPT})$,
	\begin{align}
	\label{eq.overproduction}
	\left.%\lbrace
	\begin{array}{ll}
	\Omega_{WT} & < \Omega_{WT,min} \\
	P_{cmd} & < P_{MPPT}(\Omega_{MPPT})     
	\end{array}\right\rbrace
	\rightarrow \mathrm{recovery}.
	\end{align}
	%	\begin{equation}\label{eq.overproduction}
	%	\left.%\lbrace
	%	\begin{array}{ll}
	%	\Omega_{WT} & < \Omega_{WT,min} \\
	%	P_{cmd} & < P_{MPPT}(\Omega_{MPPT})     
	%	\end{array}\right\rbrace
	%	\rightarrow \mathrm{Recovery}.
	%	\end{equation}
	\item In the {\em{recovery mode}}, the power supplied by the VSWTs ($P_{cmd}$) is based on two periods:
	following a parabolic trajectory until the middle of the rotational speed deviation ($\Omega_{V}$ in Figure~\ref{fig.control}) and through an estimated curve proportional to the difference between
	$P_{mt}(\Omega_{WT})$ and $P_{MPPT}(\Omega_{WT})$, being $x$ the proportionality constant.
	\begin{align}
	\label{eq.recov}
	\begin{array}{ll}
	P_{cmd}=a\cdot \Omega_{WT}^{2}+b\cdot \Omega_{WT}+c\cdot \Omega_{WT} & \Omega_{WT}\leq\Omega_{V} \\
	P_{cmd}=P_{MPPT}+x\cdot(P_{mt}-P_{MPPT})& \Omega_{WT}>\Omega_{V}
	%P_{cmd} & < P_{MPPT}(\Omega_{MPPT})     
	\end{array}
	\end{align}	
	The normal operation mode is recovered when either
	$\Omega_{MPPT}$ or
	$P_{MPPT}(\Omega_{MPPT})$ are reached by the VSWTs.
	\begin{align}
	\label{eq.recovery}
	\left.%\lbrace
	\begin{array}{ll}
	\Omega_{WT} \approx \Omega_{MPPT} \\
	P_{cmd} \approx P_{MPPT}(\Omega_{MPPT})     
	\end{array}\right\rbrace
	\rightarrow \mathrm{normal\;operation}.
	\end{align}
	%	\begin{equation}\label{eq.recovery}
	%	\left.%\lbrace
	%	\begin{array}{ll}
	%	\Omega_{WT} \approx \Omega_{MPPT} \\
	%	P_{cmd} \approx P_{MPPT}(\Omega_{MPPT})     
	%	\end{array}\right\rbrace
	%	\rightarrow \mathrm{Normal\;operation}.
	%	\end{equation}
\end{enumerate}

\subsection{Scenarios}\label{sec.scenarios}
Four different scenarios have been considered for simulations. The
first scenario includes only conventional generation units: 88\% comes
from thermal power plants and 12\% from hydro-power
plants. Hydro-power capacity remains constant in all the scenarios
(12\%). However, thermal and wind capacities change depending on the
scenario to be simulated by giving a power system with high
integration of RES, see Table~\ref{tab.scenarios}. The equivalent
inertia constant $H_{eq}$ determined by~\eqref{eq.Heq} is also
indicated in Table~\ref{tab.scenarios}. The power imbalance considered
is $\Delta P_{L}=0.05$~pu in all simulations.
\begin{table}[!tbp]
  \processtable{Capacity of generating units\label{tab.scenarios}}
  {\begin{tabular*}{20pc}{@{\extracolsep{\fill}}lllll@{}}\toprule
     {\bf{Source}}  & {\bf{Scenario 1}} & {\bf{Scenario 2}} & {\bf{Scenario 3}} & {\bf{Scenario 4}} \\
     \midrule
     Thermal & 88\% & 73\% & 58\% & 43\% \\ 
     Hydro-power & 12\% & 12\% & 12\% & 12\% \\ 
     Wind & 0\% & 15\% & 30\% & 45\% \\  
     $H_{eq}$ based on~\eqref{eq.Heq} &4.80~s & 4.05~s & 3.30~s & 2.55~s\\ 
     \botrule
   \end{tabular*}}{}
\end{table} 

\section{Results}\label{sec.results}

{\color{black}{According to the different methodologies discussed in
    Section~\ref{sec.inertia_estimation}, the equivalent inertia
    constant $H_{eq}$ is estimated from the frequency deviations after
    a power imbalance. Two different approaches are considered and
    compared in this work:}}
\begin{enumerate}[i. ]
\item Wind power plants without participation in frequency control.
\item Wind power plants with participation in frequency control.
\end{enumerate}

\begin{figure}[!tb]
  \centering
  \subfloat[]{\includegraphics[width = 0.99\linewidth]{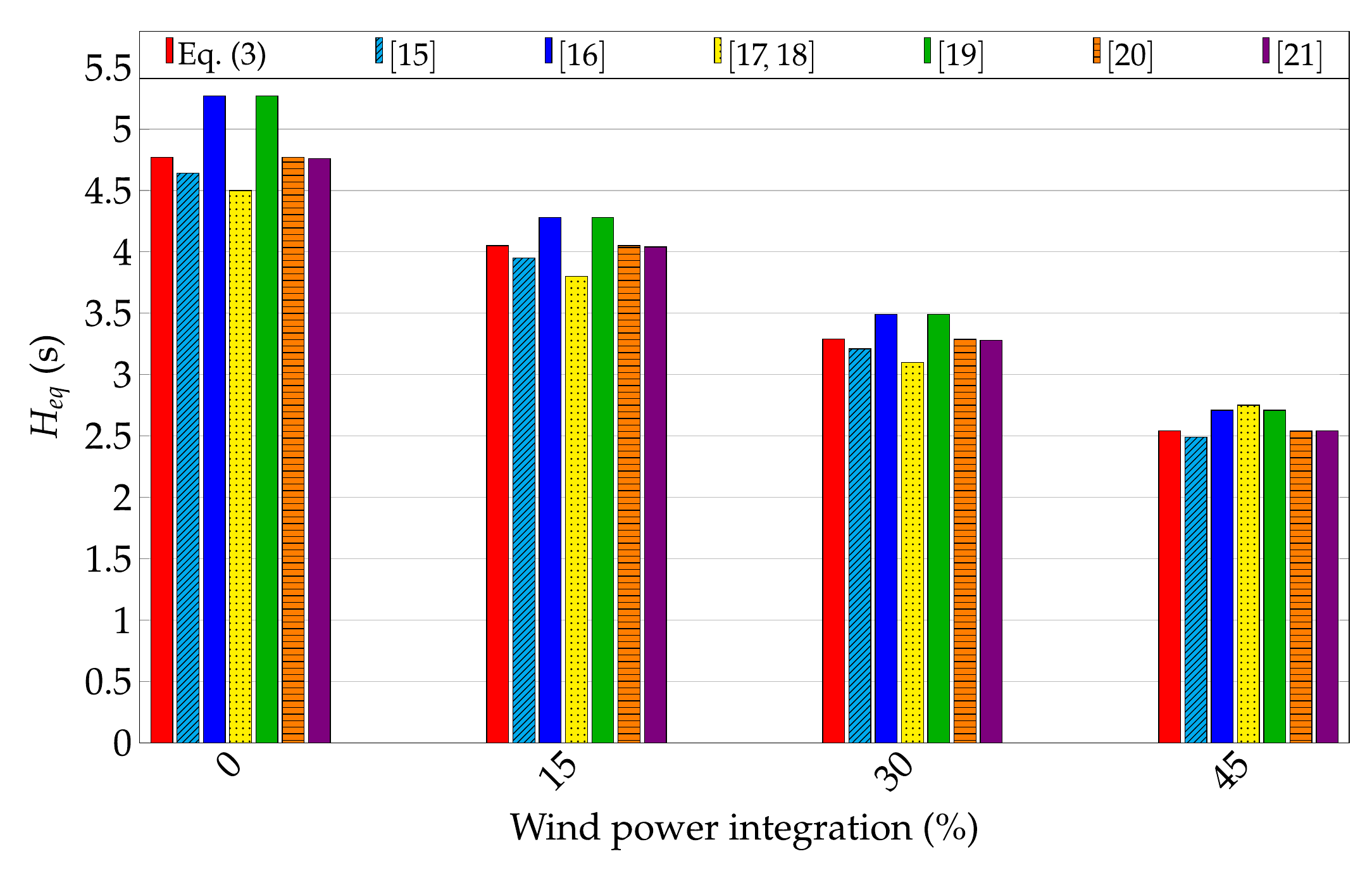}%
    \label{fig.sin_WF}} \\
  \subfloat[]{\includegraphics[width = 0.99\linewidth]{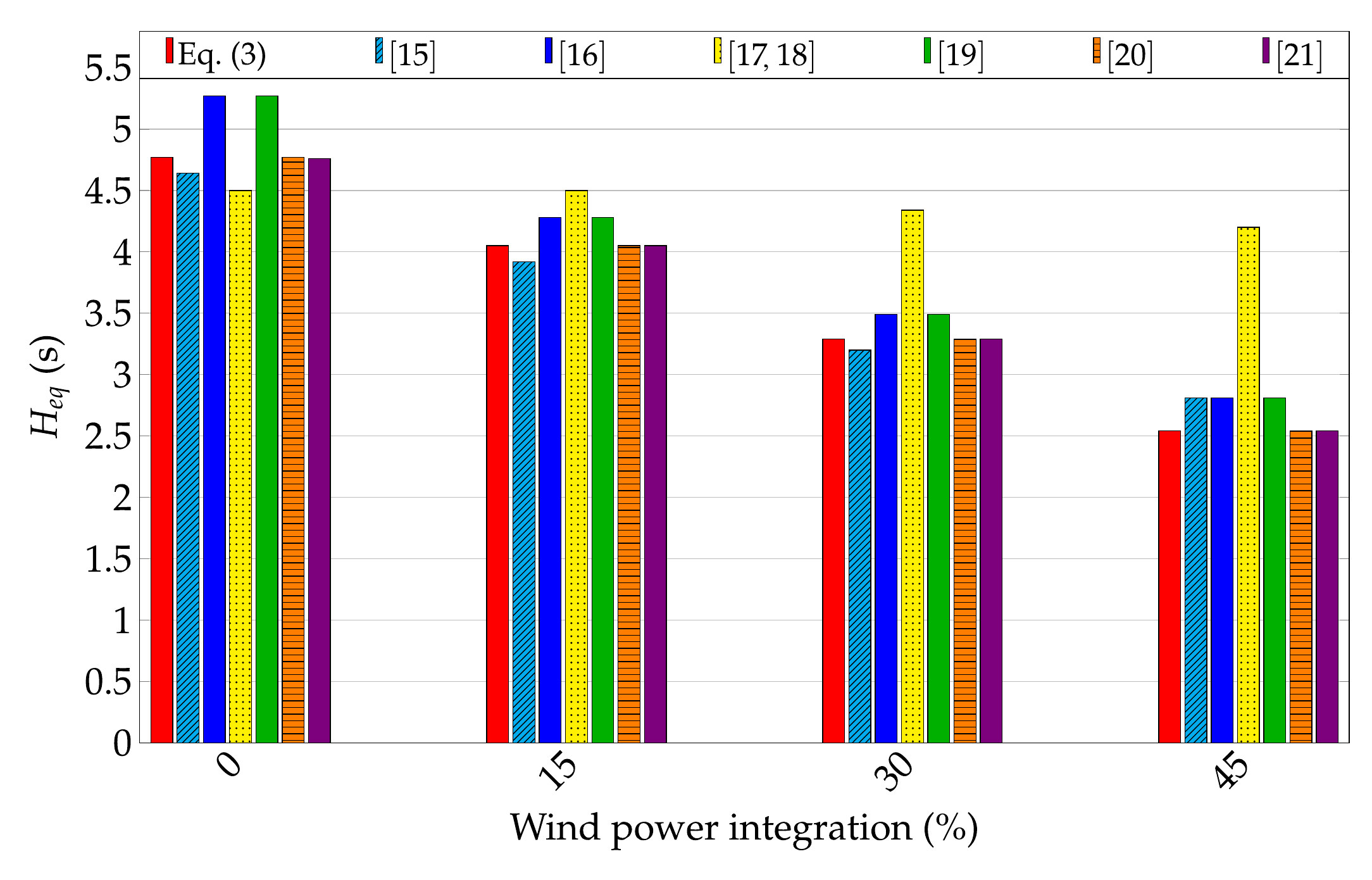}%
    \label{fig.con_WF}}
  \caption{\color{black}Comparison of equivalent inertia depending on the participation of wind power plants into frequency control}
  \figfooter{\color{black}(a)}{\color{black}Estimated $H_{eq}$ when wind power plants do not participate in frequency control}
  \figfooter{\color{black}(b)}{\color{black}Estimated $H_{eq}$ when wind power plants participate in frequency control}
  \label{fig.Heqs}
\end{figure}

%\subsection{Wind energy does not participate in frequency control}\label{sec.wind_energy_not_frequency_control}
Figure~\ref{fig.sin_WF} depicts the estimated $H_{eq}$ according to
the different methodologies without considering wind power plant
participation in frequency control. In this case,
$H_{R,eq}=H_{eq}$. The different approaches of inertia estimation
provide an accurate approximation of the directly connected rotational
inertia calculated with eq.~\eqref{eq.Heq}. The deviation from the
estimated inertia value is lower that a 10\% error.

In addition, Figure~\ref{fig.con_WF} summarizes the estimated $H_{eq}$ from the
different methodologies when wind power plants participate in
frequency control. In this case, it is expected that the estimated
$H_{eq}$ values from $\Delta f$ include the virtual inertia $H_{V,eq}$
referred to eq.~\eqref{eq.hagg2}. {\color{black}{However, as can be
    seen, most methodologies only provide the rotational inertia
    $H_{R,eq}$ directly connected to the
    grid~\cite{inoue97,chassin05,zografos17,tuttelberg18,zografos18},
    neglecting the 'virtual inertia' emulated and provided by the wind
    power plants. With these methodologies, the estimation of $H_{eq}$
    is again accurate to the value calculated by eq.~\eqref{eq.Heq},
    having a deviation lower that a 10\% error.}}

\begin{figure}[!tbp]
  \centering
  \subfloat[]{\includegraphics[width=0.4\textwidth]{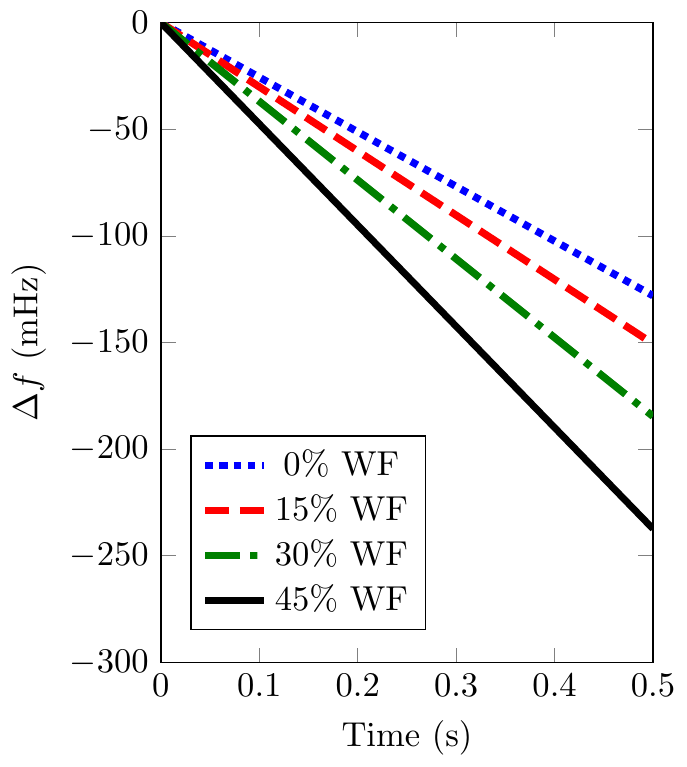}%
    \label{fig.rocof_sin}} \\
  \subfloat[]{\includegraphics[width=0.4\textwidth]{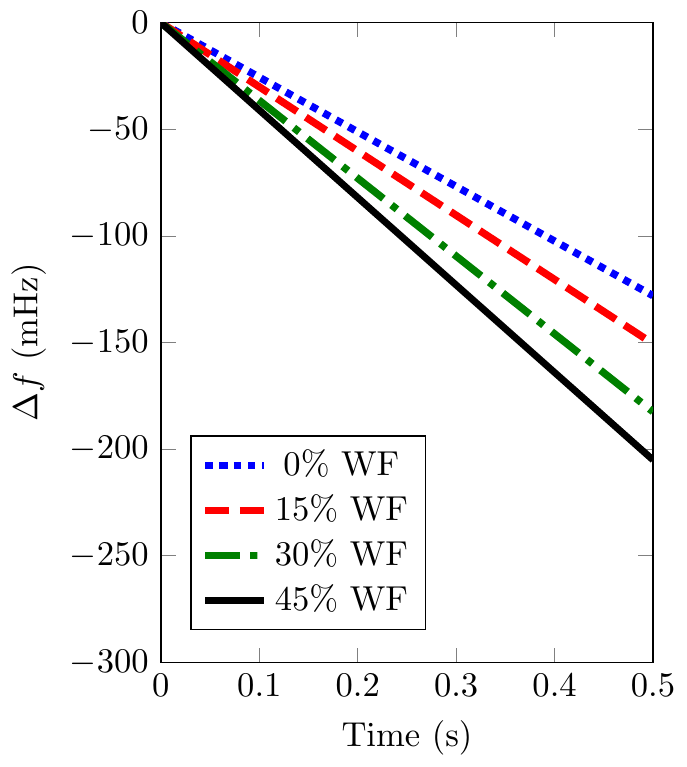}%
    \label{fig.rocof_con}}
  \caption{Comparison of ROCOF depending on the participation of wind power plants into frequency control}
  \figfooter{(a)}{Wind power plants do not participate in frequency control}
  \figfooter{(b)}{Wind power plants participate in frequency control}
  \label{fig.rocof}
\end{figure}

{\color{black}The frequency controller applied on the equivalent wind
  turbine doesn't include a derivative dependence control, see
  Section~\ref{sec.frequency_control}. As a consequence, the ROCOF is
  hardly modified in comparison to scenarios where wind power plants
  are excluded from the frequency control. At the beginning of the
  frequency oscillations, $\Delta f$ values don't change significantly
  ---see Fig. {\ref{fig.rocof}}---, regardless of the integration and
  participation of wind power plants into the frequency
  control. Table~\ref{tab.rocof} summarizes these ROCOF values
  ($mHz/s$) depending on the participation of wind power plants into
  frequency control.}

\begin{table}[!tbp]
  \processtable{{\color{black}ROCOF values ($mHz/s$) depending on the participation of wind power plants into frequency control\label{tab.rocof}}}
  {\begin{tabular*}{20pc}{@{\extracolsep{\fill}}lllll@{}}\toprule
     \multirow{2}{*}{} &  \multicolumn{4}{c}{{\bf{Wind power integration}}} \\
                       & 0\% &15\% & 30\% & 45\%\\ 
     \midrule
     Without control & -256.06 & -301.08 & -369.20 & -474.70 \\
     With control & -256.06 & -298.45 & -364.90 & -410.10 \\ 
     \botrule
   \end{tabular*}}{}
\end{table}

Methodologies~\cite{inoue97,chassin05,zografos17,zografos18} estimate
$H_{eq}$ based on the power imbalance $\Delta P$ and the
ROCOF. $\Delta P$ is the same in all the scenarios
($\Delta P=0.05$~pu), and the ROCOF values are similar regardless of
the participation of wind power plants into frequency control as
aforementioned. As a result, the estimated $H_{eq}$ barely changes
despite of including wind power plants into frequency
control. Tuttelberg~\textit{et al.} apply an impulse function to the
dynamic response, estimating $H_{eq}$ by its value at
$t=0$~\cite{tuttelberg18}. Only~\cite{wall12,wall14}, by considering
the total active power supplied and the ROCOF ---referred to
eq.~\eqref{eq.wall}---, estimates the equivalent inertia as a
combination of rotational $H_{R,eq}$ and virtual $H_{V,eq}$ inertias
as were expressed in~\eqref{eq.hagg2}.

\begin{figure}[!tb]
  \centering
  \subfloat[]{\includegraphics[width=0.5\textwidth]{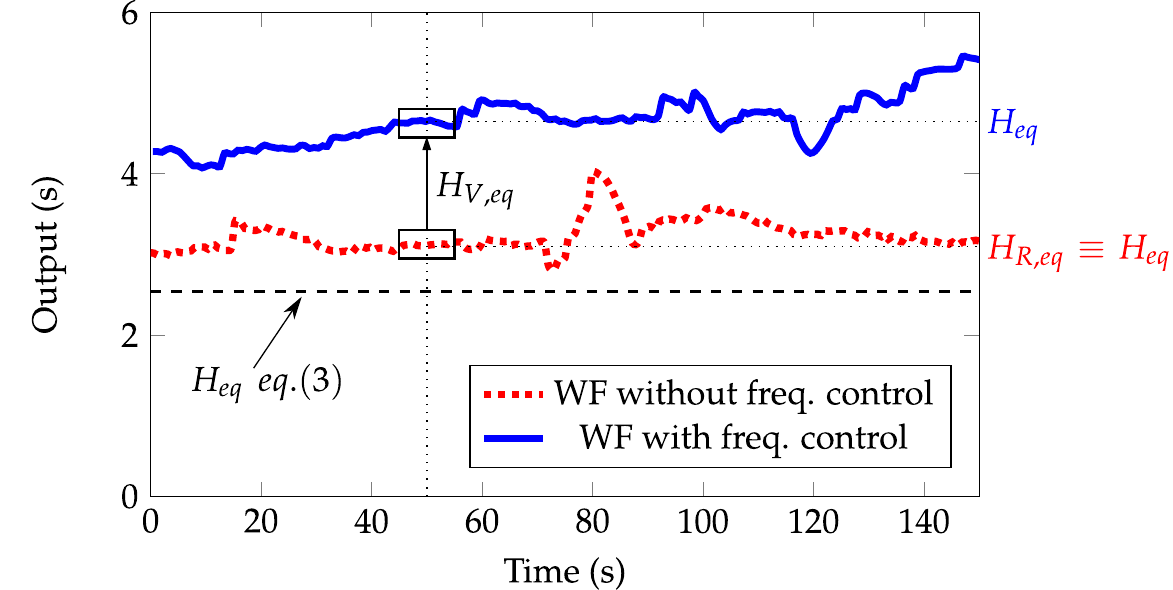}%
    \label{fig.document}}
  \\
  \subfloat[]{\includegraphics[width=0.5\textwidth]{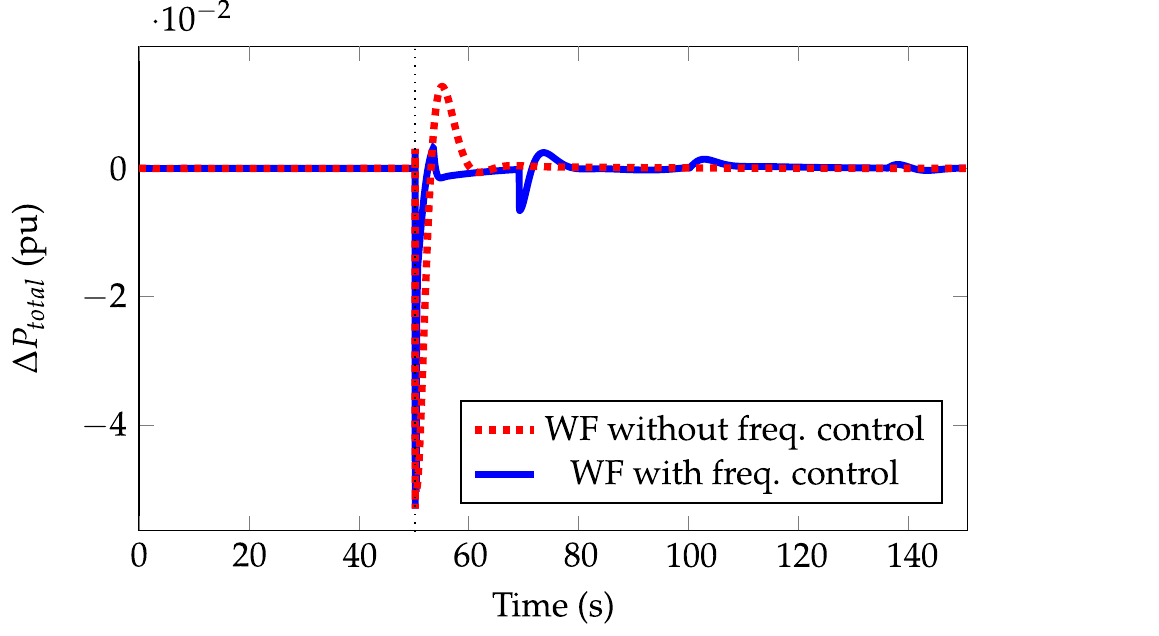}%
    \label{fig.document_ptotal}}
  \\
  \subfloat[]{\includegraphics[width=0.5\textwidth]{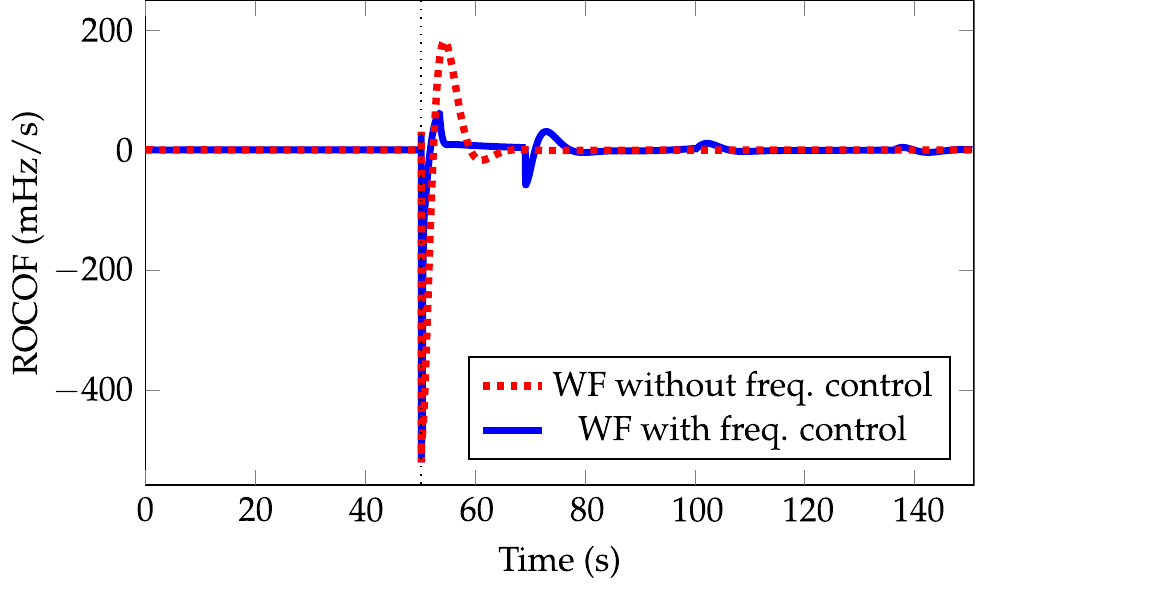}%
    \label{fig.document_rocof}}
  \caption{Estimated equivalent inertia according to~\cite{wall12,wall14}, total power variation and ROCOF in scenario 4\label{fig.h_wall}}
  \figfooter{(a)}{Estimated equivalent inertia (s)}
  \figfooter{\color{black}(b)}{\color{black}Total power variation (pu)}
  \figfooter{\color{black}(c)}{\color{black}ROCOF (mHz/s)}
\end{figure}

{\color{black}Figure~\ref{fig.h_wall} compares the equivalent inertia
  with and without frequency control from wind power plants in
  {\texttt{Scenario 4}}. This inertia is estimated according
  to~\cite{wall12,wall14}. Total power variation and ROCOF are also
  depicted for the sake of clarity. The disturbance time is
  $t_{dist}=50$~s. As indicated in~\cite{wall12,wall14} (and
  previously mentioned in Section~\ref{sec.inertia_estimation}),
  Figure~\ref{fig.document} is only the equivalent inertia around
  $t_{dist}$, as squared in the figure. Moreover, when wind power
  plants don't participate in frequency control, the equivalent
  inertia obtained is similar to the value calculated with
  eq.~\eqref{eq.Heq}, as already mentioned in
  Figure~\ref{fig.sin_WF}. However, a significant difference exists in
  the estimated equivalent inertia when wind power plants include
  frequency control. This increasing is due to the 'virtual inertia'
  provided by the wind frequency control. This virtual inertia thus
  depends on how relevant is the wind integration into the generation
  mix. Moreover, a linear relationship has been found between the wind
  power integration and the virtual inertia with $R^{2}\approx1$. The
  linear relationship can be determined as}
\begin{equation}\label{eq.linear_relationship}
H_{V,eq}=0.0357\cdot WPI\;,
\end{equation}
{\color{black}being $WPI$ the wind power integration into the grid (in \%). Considering eq.~\eqref{eq.hagg2},~\eqref{eq.linear_relationship} and the base power~$S_{B}=1350$~MW, it is obtained that the virtual inertia constant coming from wind turbines is $H_{V,WT}=3.57$~s, in line with the typical rotational inertia constants of conventional plants (refer to Section~\ref{sec.inertia_constant}) and the wind turbines inertia values proposed by some authors during the last decade~\cite{tielens12,yang12,arani13,tielens16}.}

\begin{figure}[!tbp]
	\centering
	\subfloat[]{\includegraphics[width=0.5\textwidth]{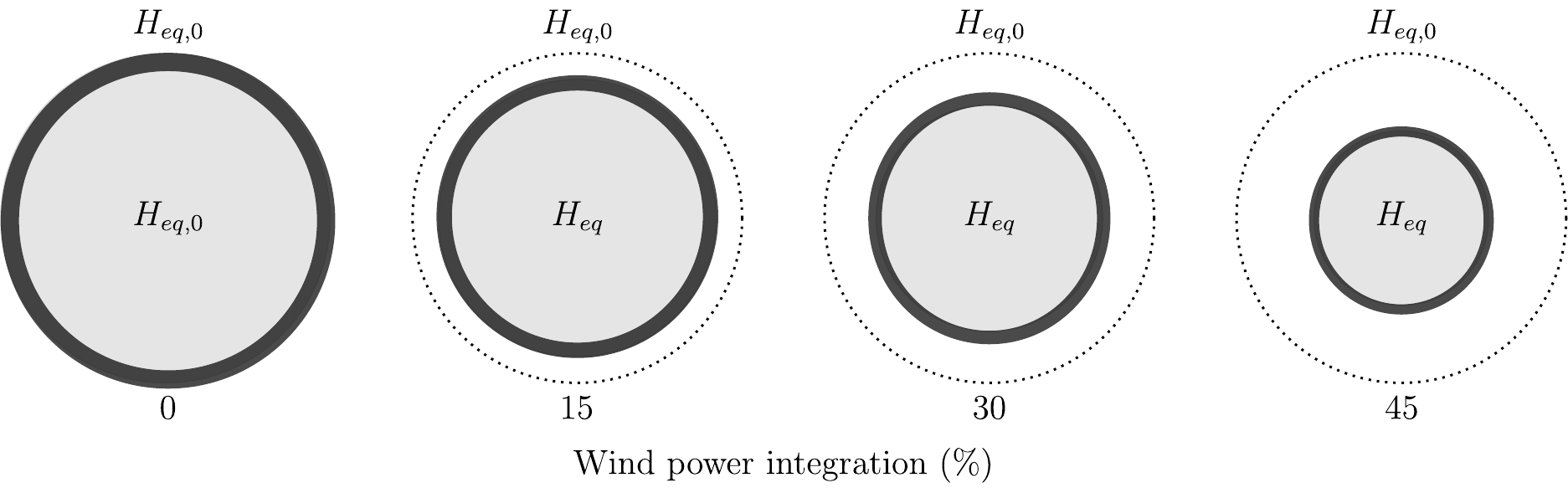}%
		\label{fig.hReq}}
	\\
	\subfloat[]{\includegraphics[width=0.5\textwidth]{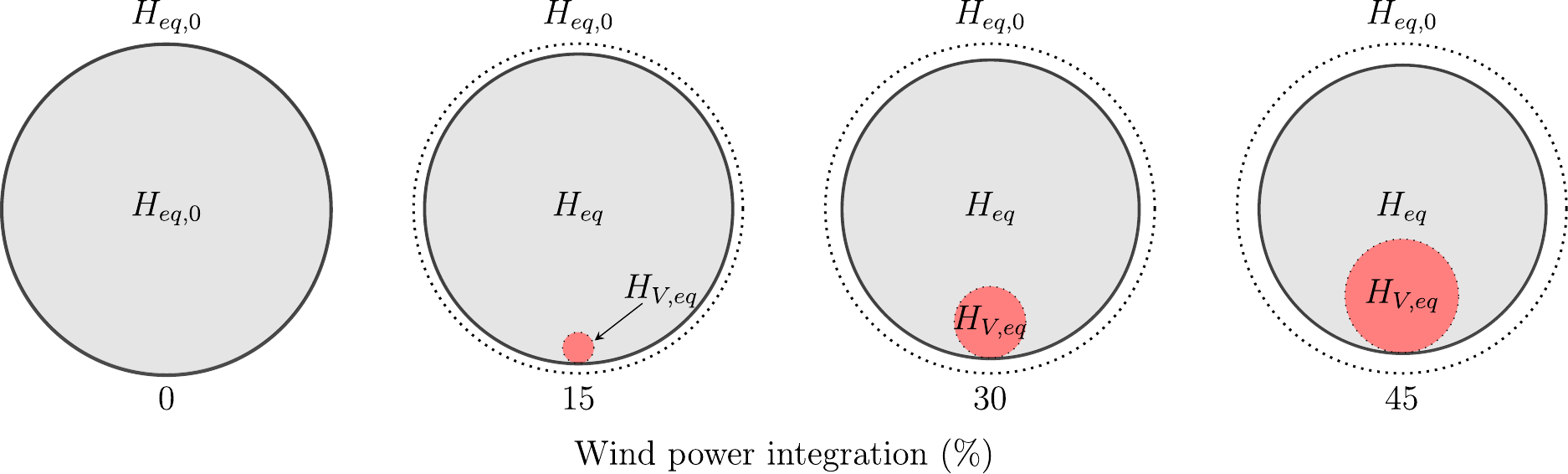}%
		\label{fig.hRhV}}
	\caption{Comparison of inertia estimation including wind power plants into frequency control}
	\figfooter{(a)}{$H_{eq}$ according to~\cite{inoue97,chassin05,zografos17,tuttelberg18,zografos18}}
	\figfooter{(b)}{$H_{eq}$ according to~\cite{wall12,wall14}}
	\label{fig.H_comparison}
\end{figure}

Finally, Figure~\ref{fig.H_comparison} summarizes the simulated
scenarios in terms of the estimated $H_{eq}$ from the different
methodologies when frequency control is also provided by wind power
plants. As can be seen, and depending on the methodology, these
approaches address significant discrepancies on the equivalent inertia
values. Actually, some of them include some virtual inertia from the
wind turbine frequency control of~\cite{fernandez18}, whereas the
others consider their effects barely significant regarding to the
equivalent system inertia. Therefore, both wind power plant frequency
control strategies and equivalent inertia estimation methodologies
must be revised in detail to give suitable results and avoid
significant discrepancies among the different proposals in the new mix
generation scenarios.

%\begin{equation}
%\begin{split}
%H_{eq}=\dfrac{\displaystyle\sum_{i=1}^{CP}H_{i}\cdot S_{B,i}+\displaystyle\sum_{j=1}^{VG}H_{V,j}\cdot S_{B,j}}{S_{B}}=\\
%=\dfrac{\displaystyle\sum_{i=1}^{CP}H_{i}\cdot S_{B,i}}{S_{B}}+\dfrac{\displaystyle\sum_{j=1}^{VG}H_{V,j}\cdot S_{B,j}}{S_{B}}=H_{R,eq}+H_{V,eq}
%\end{split}
%\end{equation}
%\begin{equation}
%H_{V,eq}=\dfrac{\displaystyle\sum_{j=1}^{VG}H_{V,j}\cdot S_{B,j}}{S_{B}}=\dfrac{H_{V,WT}\cdot S_{B,WPP}}{S_{B}}\;,
%\end{equation}

%Figure~\ref{fig.h} represents the evolution of the equivalent rotational $H_{R,eq}$ and virtual $H_{V,eq}$ inertias (in black and red, respectively), when wind power integration increases from 0 to 100\%. The green line represents the equivalent inertia $H_{eq}$, defined as the sum of rotational plus virtual inertias. Despite of the virtual inertia of wind turbines, the equivalent inertia of the system still reduces a 25\%. In any case, the value of $H_{V,eq}$ depends on the frequency control strategy implemented for wind turbines.
%\begin{figure*}[!t]
%	\centering
%	\includegraphics[width=0.5\textwidth]{figures-def/h.pdf}
%	\caption{Equivalent inertia, defined as the sum of rotational and virtual inertias}
%	\label{fig.h}
%\end{figure*}

\section{Conclusion}\label{sec.conclusion}
In this paper, an analysis and comparison of power system inertia
estimation methodologies has been carried out. Different approaches
proposed in the literature have been implemented and tested under four
different supply-side scenarios including thermal, hydro-power and
wind power plants from the supply-side, according to current mix
generation road-maps. In this way, wind power plants are increasing
their generation capacity from 15 to 45\%, reducing the thermal plants
capacity accordingly. Furthermore, wind power plants include a virtual
inertia frequency control strategy to support frequency excursions
under imbalance conditions. The inertia estimation methodologies give
an accurate value of the equivalent inertia when wind power plants do
not participate in frequency control, with a deviation error lower
than a 10\% with respect to the global rotational generation units
directly connected to grid. By including wind power plants into
frequency control, most methodologies estimate the equivalent
rotational inertia principally provided by conventional units, maintaining a
deviation error lower than 10\% in comparison with this value. One methodology estimates the equivalent inertia as a
combination of rotational and virtual inertias. The virtual
inertia constant estimated with this methodology has a value of $H_{V,WT} = 3.57$~s, in line with the
typical inertia constants of conventional plants. Therefore, wind
power plant frequency control strategies and equivalent inertia
estimation methodologies must be revised to provide consistent results
and avoid significant discrepancies among the different
alternatives. Moreover, the estimation of equivalent inertia values is
highly dependent on the wind power plant frequency control strategies,
and then, different results are determined when derivative frequency
dependence is (or not) included in the frequency strategy. Alternative
methodologies and processes should be thus proposed by the sector to
provide suitable results regarding equivalent inertia estimations in
power systems with high renewable penetration.

\section{Acknowledgments}\label{sec11}

This work is supported by the Spanish Ministry of Education, Culture and Sport ---FPU16/04282---.

\color{black}{
\section{Appendix}\label{sec.appendices}
%\appendix

\subsection{Parameters for thermal and hydro-power plants}\label{sec.params}
Table~\ref{tab.thermal} and \ref{tab.hydro} summarize the thermal and hydro-power plant parameters used in the simulations.

\begin{table}[h]
  \resizebox{\linewidth}{!}{%
  \processtable{{\color{black}Thermal power plant parameters \cite{kundur94}\label{tab.thermal}} }
  {\begin{tabular*}{20pc}{@{\extracolsep{\fill}}llll@{} }
     \toprule
     {\bf{Parameter}}  & {\bf{Description}} & {\bf{Value}} & {\bf{Units}} \\
     \midrule
     $T_{G}$ & Speed relay pilot valve & 0.20 & -- \\ 
     $F_{HP}$ & Fraction of power of high pressure section & 0.30 & -- \\ 
     $T_{RH}$ & Time constant of reheater & 7.00 & s\\ 
     $T_{CH}$ & Time constant (inlet volumes and steam chest) & 0.30 & s\\
     $R_{T}$ & Speed droop & 0.05 & pu\\ 
     $I(s)$ & Integral controller & 1.00 & --\\
     $H_{thermal}$ & Inertia constant & 5.00 & s\\
     \botrule
   \end{tabular*}}{}
 }
\end{table}
\begin{table}[h]
  \processtable{{\color{black}Hydro-power plant parameters \cite{kundur94}\label{tab.hydro}} }
  {\begin{tabular*}{20pc}{@{\extracolsep{\fill}}llll@{}}
     \toprule
     {\bf{Parameter}}  & {\bf{Description}} & {\bf{Value}} & {\bf{Units}} \\
     \midrule
     $T_{G}$ & Speed relay pilot valve & 0.20 & s \\ 
     $T_{R}$ & Reset time & 5.00 & s \\ 
     $R_{T}$ & Temporary droop & 0.38 & --\\ 
     $R_{P}$ & Permanent droop & 0.05 & --\\ 
     $T_{W}$ & Water starting time & 1.00 & s\\ 
     $R_{H}$ & Speed droop & 0.05 & pu \\
     $I(s)$ & Integral controller & 1.00 & --\\
     $H_{hydro}$ & Inertia constant & 3.00 & s\\
     \botrule
   \end{tabular*}}{}
\end{table}

\subsection{Wind turbine model}\label{sec.wpp}
The wind turbine model is based on~\cite{miller03,ullah08}. Parameters of the wind turbine model are summarized in Table \ref{tab.wt}.

\begin{table}[h]
  \processtable{{\color{black}Equivalent wind turbine parameters\cite{miller03,ullah08}\label{tab.wt}}}
  {\begin{tabular*}{20pc}{@{\extracolsep{\fill}}llll@{}}\toprule
     {\bf{Parameter}}  & {\bf{Description}} & {\bf{Value}} & {\bf{Units}} \\
     \midrule
     $v_{w}$ & Wind speed & 10.00 & m/s\\
     %$V_{WT}$ & Wind turbine voltage & 1.00 & \\
     $K_{pt}$ & Proportional constant of speed controller & 3.00 & --\\
     $K_{it}$ & Integral constant of speed controller & 0.60 & --\\
     $V_{WT}$ & Voltage of the wind turbine & 1.00 & pu \\
     $T_{con}$ & Time delay to generate the current $I_{inj}$ & 0.02 & s \\
     $T_{f}$ & Time delay to measure the active power $P_{e}$ & 5.00 & s \\
     \botrule
   \end{tabular*}}{}
\end{table}
}

%\section{References}\label{sec12}
\bibliographystyle{iet}
\bibliography{biblio}

\end{document}